\documentclass[preprint,amsmath,amssymb,nofootinbib,floatfix,aps]{revtex4}
\usepackage[pdftex]{graphicx}
\usepackage{epstopdf}
\usepackage{bm}
\usepackage{cancel}

\makeatletter
\def\Hline{
  \noalign{\ifnum0=`}\fi\hrule \@height 4.\arrayrulewidth \futurelet
  \reserved@a\@xhline}
\makeatother

\def\Journal#1#2#3#4{{#1}, {\bf #2} (#4), #3}

\def\APJ{Astrophys. J.}

\def\CompPhysCom{Comput. Phys. Commun.}
\def\EPJC{Eur. Phys. J. C}

\def\IJMPA{Int. J. Mod. Phys. A}

\def\JHEP{J. High Energy Phys.}

\def\NPB{Nucl. Phys. B}

\def\PLB{{Phys. Lett.} B}

\def\PREP{Phys. Rep.}

\def\PLBOLD{Phys. Lett.}

\def\PRL{Phys. Rev. Lett.}
\def\PRD{Phys. Rev. D}

\begin{document}

\title{Spinorial Structure of $O(3)$ and Application to Dark Matter}

\author{Teruyuki Kitabayashi}
\email{teruyuki@tokai-u.jp}

\author{Masaki Yasu\`{e}}%
\email{yasue@keyaki.cc.u-tokai.ac.jp}
\affiliation{\vspace{5mm}%
\sl Department of Physics, Tokai University,\\
4-1-1 Kitakaname, Hiratsuka, Kanagawa 259-1292, Japan\\
}

\date{\today}

\begin{abstract}
An $O(3)$ spinor, $\Phi$, as a doublet denoted by ${\bf 2}_D$ consists of an $SO(3)$ spinor, $\phi$, and its complex conjugate, $\phi^\ast$, which form $\Phi=\left(\phi,\phi^\ast\right)^T$ to be identified with a Majorana-type spinor of $O(4)$. The four gamma matrices $\Gamma_\mu$ ($\mu=1\sim 4$) are given by $\Gamma_i=\text{diag.}\left(\tau_i,\tau^\ast_i\right)$ ($i=1,2,3$) and $\Gamma_4=-\tau_2\otimes\tau_2$, where $\tau_i$ denote the Pauli matrices. The rotations and axis-reflections of $O(3)$ are, respectively, generated by $\Sigma_{ij}$ and $\Sigma_{i4}$, where $\Sigma_{\mu\nu}=[\Gamma_\mu,\Gamma_\nu]/2i$.  While $\Phi$ is regarded as a scalar, a fermionic $O(3)$ spinor is constructed out of an $SO(3)$ doublet Dirac spinor and its charge conjugate. These $O(3)$ spinors are restricted to be neutral and cannot carry the standard model quantum numbers because they contain particles and antiparticles. Our $O(3)$ spinors serve as candidates of dark matter. The $O(3)$ symmetry in particle physics is visible when the invariance of interactions is considered by explicitly including their complex conjugates.  It is possible to introduce a dark gauge symmetry based on $SO(3)\times\boldsymbol{Z}_2$ equivalent to $O(3)$, where the $\boldsymbol{Z}_2$ parity is described by a $U(1)$ charge giving 1 for a particle and $-1$ for an antiparticle.  The $SO(3)$ and $U(1)$ gauge bosons turn out to transform as the axial vector of $O(3)$ and the pseudoscalar of $O(3)$, respectively.  This property is related to the consistent definition of the nonabelian field strength tensor of $O(3)$ or of the U(1) charge of the O(3)-transformed spinor.  To see the feasibility of our dark matter models, we discuss scalar dark matter phenomenology based on the dark $U(1)$ gauge model.
\end{abstract}

\maketitle



\section{Introduction}
The cosmological observation of dark matter \cite{Planck2016} has inspired theoretical interest in seeking possible physics of dark matter \cite{Arkani-Hamed2009}. However, theoretical description of dark matter to date remains unclear and a candidate of dark matter is not provided by the standard model. A simple candidate exhibits the property that dark matter does not couple to the standard model particles and is present in the so-called dark sector \cite{Alexander2016}. The dark sector (or hidden sector) may contain new particles that couple only indirectly to ordinary matter. These new particles are expected to have masses well below the weak-scale. The dark sector particles communicate with the standard model particles via \lq\lq dark matter portals\rq\rq.  The typical dark matter portal includes the coupling of Higgs particle to dark scalars (as a Higgs portal), of flavor neutrinos to sterile neutrinos (as a neutrino portal) and of photons to dark photons (as a vector portal) \cite{Arcadi2018}.

The key issue to discuss is how to realize the constraint on the dark sector particles that they do not have direct couplings to ordinary matter. The appropriate invariance of the dark sector based on certain symmetries may forbid the dark sector particles to couple to ordinary matter. Discrete symmetries such as $Z_2$ and a $U(1)$ symmetry are the simpler candidates that also ensure stability of dark matter although the origin of the symmetries is not naturally understood except for the need to constrain the dark sector.

The constraint on the dark sector is naturally satisfied if dark matter transforms as the recently advocated spinorial doublet of the $O(3)$ symmetry \cite{Kitabayashi2018}. We have presented an irreducible representation ${\bf 2}_D$ as the spinorial doublet of $O(3)$, which is applied to models of quarks and leptons possessing the discrete $S_4$ symmetry that has been considered as a promising flavor symmetry of quarks and leptons \cite{Pakvasa1979,Derman1979,Lam2008PRL,Bernigaud2018}. Since ${\bf 2}_D\otimes{\bf 2}_D={\bf 1}\oplus{\bf 3}$, the $O(3)$ spinor provides $S_4$-invariant couplings to the standard model particles.  Since $S_4$ is known as a subgroup of $O(3)$, the $S_4$ spinor can be based on a four component $O(3)$ spinor, which is composed of the two component $SO(3)$ spinor and its complex conjugate.  A remarkable feature is that the $O(3)$ spinor as an elementary particle contains a particle as the $SO(3)$ spinor and an antiparticle as its conjugate in the same multiplet of ${\bf 2}_D$ so that our $O(3)$ spinors cannot have no quantum numbers of the standard model. It is clear that the definition of the $O(3)$ spinor itself forbids dark sector particles to directly couple to ordinary matter if dark matter consists of the $O(3)$ spinor.

In this paper, we would like to enlarge our previous argument \cite{Kitabayashi2018} on the possible existence of an $O(3)$ spinor on the basis of the four component spinor of the $O(4)$ symmetry and to include the physical aspect of the $O(3)$ spinor.  Since particle physics involves fermionic degrees of freedom, our $O(3)$ spinors must include a fermionic $O(3)$ spinor.  To introduce the fermionic $O(3)$ spinor needs careful examination because it is simultaneously the Dirac spinor.  As expected, it is understood that the charge conjugate is employed instead of the complex conjugate to be consistent with the Dirac spinor.  We have to confirm that such a fermionic spinor correctly transforms under the $O(3)$ transformation so that the internal $O(3)$ symmetry becomes orthogonal to the Lorentz symmetry.

It is further demonstrated that a gauged $O(3)$ symmetry is based on the mathematical equivalence of $O(3)$ to $SO(3) \times \boldsymbol{Z}_2$, where $\boldsymbol{Z}_2$ is generated by the parity inversion.  For the $O(3)$ spinor, it is found that $\boldsymbol{Z}_2$ is described by a spinorial $\boldsymbol{Z}_2$ parity operator, whose eigenvalues are $1$ for the particle as the $SO(3)$ spinor and $-1$ for the antiparticle as its conjugate that does imply the appearance of a $U(1)$ symmetry.  The typical feature of the gauged $O(3)$ symmetry is that the gauge bosons transform as the axial vector of $O(3)$ (or the pseudoscalar of $O(3)$).  This property is related to the consistent definition of the nonabelian field strength tensor (or of the $U(1)$ charge of the $O(3)$-transformed spinor).

The present article is organized as follows: In Sec. \ref{sec:O3-spinor}, we clarify spinorial structure of $O(3)$ based on the $O(4)$ symmetry.  In Sec.\ref{sec:particlephysics}, we discuss application of the $O(3)$ spinor to particle physics. For a scalar and a fermion taken as the $O(3)$ spinor, we construct $O(3)$-invariant lagrangians.  The consistency with the Dirac spinor is clarified.  Also discussed is the possible inclusion of the gauged $O(3)$ symmetry based on $SO(3) \times \boldsymbol{Z}_2$.  Section \ref{sec:darksector} deals with discussions on our candidates of dark matter and dark gauge bosons.  A variety of implementation of dark gauge symmetries based on $O(3)$ is discussed.  In the next section, Sec.\ref{sec:phenomenology}, to see the feasibility of our scenarios, we present dark matter phenomenology based on the dark $U(1)$ gauge model. The final section Sec.\ref{sec:summary} is devoted to summary and discussions. 

\section{$O(3)$ Spinor}\label{sec:O3-spinor}
\subsection{Manifestation of Spinorial Structure}\label{sec:BriefReviewSpinor}
We have advocated the use of the $O(3)$ spinor doublet as an $S_4$ spinor representation, which is composed of an $SO(3)$ spinor $\phi$:
\begin{equation}
\phi=\left(
  \begin{array}{c}
     \phi_1   \\
     \phi_2  \\
  \end{array}
\right),
\end{equation}
from which $\Phi$ as the $O(3)$ spinor is defined to be:
\begin{equation}
\Phi=\left(
  \begin{array}{c}
     \phi   \\
     \phi^\ast  \\
  \end{array}
\right),
\label{Eq:DefOfPhi}
\end{equation}
as a four component spinor, transforming as ${\bf 2}_D$ \cite{Kitabayashi2018}. Since $O(3)$ is contained in the $O(4)$ symmetry, which enables us to discuss the parity inversion, we first introduce the standard form of the gamma matrices $\gamma_\mu$ ($\mu=1\sim 4$):
\begin{equation}
{\gamma _i} = \left( {\begin{array}{*{20}{c}}
  {{\tau _i}}&0 \\ 
  0&{ - {\tau _i}} 
\end{array}} \right)
= {\tau _i} \otimes {\tau _3},
\quad
{\gamma _4} = \left( {\begin{array}{*{20}{c}}
  0&I \\ 
  I&0 
\end{array}} \right)
=
I \otimes {\tau _1},
\end{equation}
where $\tau_i$ $(i=1,2,3)$ stand for the three Pauli matrices. The action of generators $\gamma_{ij} = [\gamma_i,\gamma_j]/2i$ on $\Phi$ must be compatible with the definition of $\Phi=(\phi,\phi^\ast)^T$.  It is known that this kind of $\Phi$ takes the form of ($\phi,i\tau_2\phi^\ast)^T$ as the proper spinor of $O(4)$, which can be regarded as a Majorana representation of the $O(4)$ spinor. For $(\phi,\phi^\ast)$, it is straightforward to find $\Gamma_\mu$ from $\gamma_\mu$ by $W=\text{diag.}(I,i\tau_2)$:
\begin{equation}
{\Gamma _i} =
{W^\dag }{\gamma _i}W
=
\left( {\begin{array}{*{20}{c}}
  {{\tau _i}}&0 \\ 
  0&{\tau _i^\ast } 
\end{array}} \right),
\quad
{\Gamma _4} =
{W^\dag }{\gamma _4}W
=
 \left( {\begin{array}{*{20}{c}}
  0&{ i{\tau _2}} \\ 
  {-i{\tau _2}}&0 
\end{array}} \right)
=
- {\tau _2} \otimes {\tau _2},
\end{equation}
leading to the generators $\Sigma_{\mu\nu} = [\Gamma_\mu,\Gamma_\nu]/2i$ of $O(4)$ acting on $\Phi$. 

The $i$-th axis reflection of $O(3)$ on $\Phi^\dagger\Gamma_i\Phi$ as the vector of $O(3)$ is induced by $\Sigma_{i4}$, which are explicitly written as
\begin{equation}
  {\Sigma_{14}} = \left( {\begin{array}{*{20}{c}}
  0&{i{\tau _3}} \\ 
  { - i{\tau _3}}&0 
\end{array}} \right),  
\quad
{\Sigma_{24}} = \left( {\begin{array}{*{20}{c}}
  0&I \\ 
  I&0 
\end{array}} \right),
\quad
{\Sigma_{34}} = \left( {\begin{array}{*{20}{c}}
  0&{ - i{\tau _1}} \\ 
  {i{\tau _1}}&0 \\
\end{array}} \right).
\end{equation}
Altogether, we obtain the following rotation matrices of $O(3)$ denoted by $D_{ij}(\sigma,\theta)$ for $\sigma=1$ and $\sigma=-1$, respectively, taking care of the $SO(3)$ rotations and those including the parity inversion:
\begin{equation}
D_{ij}\left(1,\theta\right)= \exp\left(-i\frac{\Sigma_{ij}}{2}\theta\right),
\quad
D_{ij}\left(-1,\theta\right)= \exp\left(-i\frac{\Sigma_{ij}}{2}\theta\right)\Sigma_{i4}.
\label{Eq:Sij-sigma}
\end{equation}
In terms of the $SO(3)$ rotations on $\phi$ denoted by $S_{ij}(1,\theta)$ = $\exp(-i\tau_{ij}\theta/2)$, where $\tau_{ij}=[\tau_i,\tau_j]/2i$, $D_{ij}(\sigma,\theta)$ are expressed as follows \cite{Kitabayashi2018}:
\begin{eqnarray}
D_{ij}(1,\theta)=\left(
  \begin{array}{cc}
     S_{ij}(1,\theta) & 0 \\
     0 & S_{ij}^\ast (1,\theta)  \\
  \end{array}
\right),
\label{Eq:DijOrg}
\end{eqnarray}
and
\begin{eqnarray}
D_{ij}(-1, \theta)&=&\left(
  \begin{array}{cc}
     0&  S_{ij}(-1,\theta)  \\
     S^\ast_{ij}(-1,\theta)&0  \\
  \end{array}
\right),
\label{Eq:DijOrg2}
\end{eqnarray}
where
\begin{equation}
S_{12}(-1,\theta)= iS_{12}(1,\theta)\tau_3,
\quad
S_{23}(-1,\theta)= S_{23}(1,\theta),
\quad
S_{31}(-1,\theta)= -iS_{31}(1,\theta)\tau_1.
\label{Eq:S122331-minus-sigma}
\end{equation}

The definition of $\Phi$ given by Eq.(\ref{Eq:DefOfPhi}) can be generalized to include a parameter $\eta$, which commutes with $S_{ij}(1,\theta)$. The generalized form of the $O(3)$ spinor takes the form of
\begin{equation}
\Phi=\left(
  \begin{array}{c}
     \phi   \\
     \eta\phi^\ast  \\
  \end{array}
\right),
\label{Eq:DefOfPhiGeneral}
\end{equation}
for $\eta^\ast\eta=I$, or equivalently,
\begin{equation}
\Phi_{a+2} = \eta\Phi^\ast_a,
\label{Eq:ConsistencyCondGeneral}
\end{equation}
for $a=1,2$, as the consistency condition on $\Phi$, which will be used to find the fermionic $O(3)$ spinor.  The action of $D_{ij}(-1, \theta)$ on $\Phi$ yields 
\begin{eqnarray}
\Phi^\prime_a=\sum\limits_{b = 1}^2S_{ij}(-1,  \theta)_{ab}\Phi_{b+2},
\label{Eq:PhiPrime}\\
\Phi^\prime_{a+2}=\sum\limits_{b = 1}^2 S^\ast_{ij}(-1, \theta)_{ab}\Phi_b.
\label{Eq:PhiPrimeAst}
\end{eqnarray}
The consistency condition ensures that the complex conjugate of Eq.(\ref{Eq:PhiPrime}) coincides with Eq.(\ref{Eq:PhiPrimeAst}).  For $D_{ij}(1, \theta)$, the condition is obviously satisfied.  The simplest choice of $\eta=I$ giving
\begin{equation}
\Phi_{a+2} = \Phi^\ast_a,
\label{Eq:ConsistencyCond}
\end{equation}
corresponds to Eq.(\ref{Eq:DefOfPhi}).

\subsection{Parity inversion}\label{sec:ParityInversion}
Although $\Sigma_{i4}$ is the axis-inversion operator, it is instructive to cast $\Sigma_{i4}$ into the following form: 
\begin{equation}
{\Sigma_{i4}} ={\Sigma _i}{\tilde P}
\quad
\text{for}~
{\tilde P} =\left( {\begin{array}{*{20}{c}}
  0&{{\tau _2}} \\ 
  {{\tau _2}}&0 \\
\end{array}} \right),
\label{Eq:SigmaTildawithP}
\end{equation}
where $\Sigma_i=\epsilon_{ijk}\Sigma_{jk}$ given by
\begin{equation}
{\Sigma _i} = \left( {\begin{array}{*{20}{c}}
  {{\tau _i}}&0 \\ 
  0&{-\tau _i^\ast } 
\end{array}} \right),
\end{equation}
To see the r\^{o}les of $\Sigma _i$ and $\tilde P$, we discuss the case of $\Sigma_{14}$ for $D_{12}\left(-1,\theta\right)$.  Noticing that $\Sigma_1 = iD_{23}(1,\pi)$, we obtain that ${D_{12}}( { - 1,\theta })$ = ${D_{12}}( { 1,\theta }){D_{23}}( {1,\pi })i\tilde P$, which suggests that
\begin{itemize}
	\item $\Sigma_1 = iD_{23}(1,\pi)$  describes the reflection in the vector space;
	\item ${\tilde P}$ describes the reflection in the spinor space as the parity operator, leading to the interchange of $\phi$ and $\phi^\ast$;
\end{itemize}
and similarly for others.  As a result, we obtain that
\begin{eqnarray}
{D_{12}}\left( { - 1,\theta } \right) 
&=&
{D_{12}}\left( { 1,\theta } \right){D_{23}}\left( {1,\pi } \right)i\tilde P,
\quad
{D_{23}}\left( { - 1,\theta } \right) 
=
{D_{23}}\left( { 1,\theta } \right){D_{31}}\left( {1,\pi } \right)i\tilde P,
\nonumber\\
{D_{31}}\left( { - 1,\theta } \right) 
&=&
{D_{31}}\left( { 1,\theta } \right){D_{12}}\left( {1,\pi } \right)i\tilde P,
\label{Eq:OthreeParity}
\end{eqnarray}
which reflect that $O(3)=SO(3)\times\boldsymbol{Z}_2$.  

This decomposition of Eq.(\ref{Eq:OthreeParity}) is equivalent to the one for $T_{ij}(\sigma, \theta)$ (=$-T_{ji}(\sigma, \theta)$) describing the corresponding transformation matrices acting on the $O(3)$ vector space:
\begin{eqnarray}
{T_{12}}\left( { - 1,\theta } \right) &=& {T_{12}}\left( { 1,\theta } \right)T_{23}\left( {1,\pi } \right)P^{(3)},
\quad
{T_{23}}\left( { - 1,\theta } \right)  =  {T_{23}}\left( { 1,\theta } \right)T_{31}\left( {1,\pi } \right)P^{(3)},
\nonumber\\
{T_{31}}\left( { - 1,\theta } \right) &=& {T_{31}}\left( { 1,\theta } \right)T_{12}\left( {1,\pi } \right)P^{(3)},
\label{Eq:Inversion}
\end{eqnarray}
where $P^{(3)}$ denotes the parity operator given by $P^{(3)}=\text{diag.}(-1,-1,-1)$ and
\begin{eqnarray}
T_{12}(\sigma,\theta)
& = &
 \left(
  \begin{array}{ccc}
    \sigma \cos \theta & - \sin \theta & 0 \\
    \sigma \sin \theta  & \cos\theta & 0  \\
    0 & 0 & 1\\
  \end{array}
  \right),
\quad
T_{23}(\sigma,\theta)
=
  \left(
  \begin{array}{ccc}
    1 & 0 & 0\\
    0 & \sigma \cos \theta & -\sin \theta  \\
    0 & \sigma \sin \theta  & \cos\theta   \\
  \end{array}
  \right),
\nonumber \\
T_{31}(\sigma,\theta)
& = &
   \left(
  \begin{array}{ccc}
    \cos \theta & 0 &  \sigma \sin \theta  \\
    0 & 1 & 0\\
    -\sin \theta  & 0 & \sigma  \cos\theta  \\
  \end{array}
  \right).
  \label{Eq:T12T23T13}
\end{eqnarray}
The operator $\tilde P$ corresponding to $P^{(3)}$ plays a r\^{o}le of the spinorial parity operator, whose eigenvalues are $(1,1,-1,-1)$.  As a result, $\tilde P$ is equivalent to the following operator $P$:
\begin{equation}
P =  - i{\Gamma _1}{\Gamma _2}{\Gamma _3}
=\left(
  \begin{array}{cc}
    I & 0   \\
     0  & -I\\
  \end{array}
\right),
\label{Eq:ParityOperator}
\end{equation}
which is responsible for the $\boldsymbol{Z}_2$ parity in the spinorial space. As a result, $\phi$ and $\phi^\ast$ are distinguished by the parity operator, implying the appearance of a $U(1)$ symmetry associated with $P$.  Note that there is a relation of 
\begin{equation}
\Sigma_i=P\Gamma_i=\Gamma_iP.
\label{Eq:SigmabyGamma}
\end{equation}
%

\subsection{G-Parity}\label{sec:GParity}
We introduce a G-conjugate state, $\Phi^G$, which corresponds to a complex conjugate transforming as ${\bf 2}_D$. The G-conjugate state can be defined by
\begin{equation}
\Phi^G = G\Phi^\ast,
\label{Gparity}
\end{equation}
where
\begin{equation}
G = i\Sigma_2.
\end{equation}
Instead of Eq.(\ref{Gparity}), we also find that
\begin{equation}
\Phi^G = i{\tilde P}\Phi.
\label{GparityFromP}
\end{equation}
This spinor doublet is transformed by $D_{ij}^G\left( { \sigma,\theta } \right)$:
\begin{equation}
D_{ij}^G\left( { \sigma,\theta } \right) = GD_{ij}^\ast\left( { \sigma,\theta } \right)G^{ - 1},
\end{equation}
which should be equal to $D_{ij}\left( { \sigma,\theta } \right)$ up to phases so that $\Phi^G$ transforms as ${\bf 2}_D$.  For Eq.(\ref{Eq:Sij-sigma}), $D_{ij}^G\left( { \sigma,\theta } \right)$ = $D_{ij}\left( { \sigma,\theta } \right)$ is satisfied.

\subsection{Axial vector and pseudoscalar of $O(3)$}\label{sec:AxialPseudo}
Since $\Gamma_\mu$ is the vector of $O(4)$, $\Gamma_i$ is the $O(3)$ vector. By the action of $\tilde P$ as the spinorial version of $P^{(3)}$, we have the following results, 
\begin{equation}
{{\tilde P}^\dag }\Gamma_i\tilde P =  -\Gamma_i,
\quad
{{\tilde P}^\dag }\tilde\Gamma_i\tilde P =  -\tilde\Gamma_i.
\quad
{{\tilde P}^\dag }\Sigma_i\tilde P =  \Sigma_i,
\quad
{{\tilde P}^\dag }\tilde\Sigma_i\tilde P =  \tilde\Sigma_i.
\quad
{{\tilde P}^\dag }P\tilde P =  -P,
\label{Eq:Rotation}
\end{equation}
where $\tilde\Sigma_i=\Sigma_{i4}$ and $\tilde\Gamma_i=P\tilde\Sigma_i$. These relations correctly describe the property of the parity inversion that the vector changes its sign but the axial vector does not change its sign and that the pseudoscalar changes its sign.  Therefore, we obtain that
\begin{itemize}
	\item $\vec{\Gamma}$ and $\vec{\tilde\Gamma}$ are the vectors of $O(3)$;
	\item $\vec{\Sigma}$ and $\vec{\tilde\Sigma}$ are the axial vectors of $O(3)$;
	\item $P$ is the pseudoscalars of $O(3)$,
\end{itemize}
where 
\begin{equation}
\vec{\Gamma}= {\Gamma_1}\vec{i} + {\Gamma_2}\vec{j} + {\Gamma_3}\vec{k},
\end{equation}
and similarly for $\vec\Sigma$, $\vec{\tilde\Gamma}$ and $\vec{\tilde\Sigma}$.  The rotations generated by $D_{ij}(\sigma,\theta)$ provide the correct description of the spinor-rotations because $D_{ij}(\sigma,\theta)$ turn out to give
\begin{eqnarray}
D_{ij}^\dag \left( { \sigma,\theta } \right)\Gamma_m{D_{ij}}\left( { \sigma,\theta } \right) &=& T_{ij}\left(\sigma,\theta\right)_{mn}\Gamma_n,
\nonumber\\
D_{ij}^\dag \left( { \sigma,\theta } \right)\Sigma_m{D_{ij}}\left( { \sigma,\theta } \right) &=& \sigma T_{ij}\left(\sigma,\theta\right)_{mn}\Sigma_n,
\nonumber\\
D_{ij}^\dag \left( { \sigma,\theta } \right)P{D_{ij}}\left( { \sigma,\theta } \right) &=& \sigma P.
\label{Eq:GammaSigmaPRotation}
\end{eqnarray}
These transformation properties are consistent with those indicated by Eq.(\ref{Eq:Rotation}).  For the remaining $SO(3)$ vectors, $\vec{\tilde \Gamma}$ and $\vec{\tilde \Sigma}$, since there are relations of $\vec{\tilde \Gamma}\Phi$ = $- i\vec{\Gamma}\Phi ^G$ and $\vec{\tilde \Sigma}\Phi$ = $- i\vec{\Sigma}\Phi ^G$, the action on $\vec{\tilde \Gamma}$ and $\vec{\tilde \Sigma}$ can be understood from that on $\vec\Gamma$ and $\vec\Sigma$.

We note that the property of $P$ being the pseudoscalar operator provides us an important observation on the transformed state of $\Phi$ given by $\Phi^\prime$=$D_{ij}(-1, \theta)\Phi$.  The eigenvalues of $P$ for $\Phi^\prime$ are calculated by
\begin{equation}
P\Phi^\prime=D_{ij}(-1, \theta)\left(-P\Phi\right),
\label{Eq:PseudoP}
\end{equation}
from which we find that $-P\Phi$ is transformed into $P\Phi^\prime$.  The eigenvalues of $P$ for $\Phi^\prime$ are opposite in signs to $\Phi$.  In terms of $\phi$, $\phi^\prime = S_{ij}(-1,\theta)\phi^\ast$ given by $D_{ij}(-1,\theta)$ indicates that $P$ for $\phi^\prime$ is equal to $P$ for $\phi^\ast$.  In this pseudoscalar case, $\phi^\prime = S_{ij}(-1,\theta)\phi^\ast$ and its complex conjugate are consistent with $P\Phi^\prime$=$D_{ij}(-1, \theta)(-P\Phi)$.  The parity operator $P$ being the pseudoscalar is a key ingredient to later introduce a $U(1)$ symmetry into the $O(3)$ spinor. 

\section{Particle Physics}\label{sec:particlephysics}
To see the feasibility of the $O(3)$ spinors in particle physics, we construct $O(3)$-invariant lagrangians for a bosonic and fermionic $O(3)$ spinor.  It is found that the appearance of the $O(3)$ symmetry in particle physic can be rephrased as \lq\lq the $O(3)$ symmetry gets visible when we consider the invariance of interactions by including their complex conjugates\rq\rq. For example, the $SU(2)$ gauge interaction with an $SU(2)$-doublet scalar (as the $SO(3)$ spinor), $\phi$, contains $ig\tau_i$ with $g$ as the gauge coupling of the $SU(2)$ gauge boson, which is equivalent to $-ig\tau^\ast_i$ for $\phi^\ast$. When $\phi$ is combined with $\phi^\ast$ to form $\Phi$, the gauge interaction contains $\text{diag.}(\tau_i,-\tau^\ast_i)$ for $\Phi$, which is nothing but $\Sigma_i$.  Since $\Sigma_i$ are the generators of $O(3)$, the $O(3)$ symmetry can be gauged to include the $O(3)$ gauge bosons as our elementary particles.

\subsection{Scalar}\label{sec:scalar}
Let us start by utilizing $\Phi$ as a scalar. As already demonstrated, $\Phi^\dagger\Phi$ and $\Phi^\dagger\Gamma_i\Phi$ behave as {\bf 1} and {\bf 3}, respectively.  Because of ${\tilde X}_i\Phi$ = $- iX_i\Phi ^G$, where ${\tilde X}_i$ represents ${\tilde \Gamma}_i$ and ${\tilde \Sigma}_i$, bilinears containing ${\tilde \Gamma}_i$ or ${\tilde \Sigma}_i$ are equivalent to those containing $\Gamma_i$ or $\Sigma_i$. The relevant bilinears are $\Phi^\dagger P\Phi$, $\Phi^\dagger\Sigma_i\Phi$, $\Phi^{G \dagger}\Phi$, $\Phi^{G \dagger} P\Phi$, $\Phi^{G \dagger}\Gamma_i\Phi$ and $\Phi^{G \dagger}\Sigma_i\Phi$. The computations are straightforward to obtain that %
\begin{enumerate}
	\item $\Phi^\dagger\Phi$(=$\Phi^{G\dagger}\Phi^G$)=${\phi ^\dag }\phi  + {\phi ^T}{\phi ^\ast }$=2${\phi ^\dag }\phi$;
	\item $\Phi^\dagger P\Phi$(=$\Phi^{G\dagger} P\Phi^G$)=${\phi ^\dag }\phi  - {\phi ^T}{\phi ^\ast }$=0;
	\item $\Phi^\dagger \Gamma_i\Phi$(=$-{\Phi ^{G\dag }}{\Gamma _i}{\Phi ^G}$)=${\phi ^\dag }{\tau _i}\phi  + {\phi ^T}\tau _i^ \ast {\phi ^ \ast }$=2${\phi ^\dag }{\tau _i}\phi$;
	\item $\Phi^\dagger \Sigma_i\Phi$(=$-\Phi^{G\dagger} \Sigma_i\Phi^G$)=${\phi ^\dag }{\tau _i}\phi  - {\phi ^T}\tau _i^ \ast {\phi ^ \ast }$=0;
	\item $\Phi^{G \dagger}\Phi$=$ - i\left( {{\phi ^T}{\tau _2}\phi  - {\phi ^\dag }\tau _2^\ast {\phi ^\ast }} \right)$=0, where ${\tau _2} = -\tau^T_2$;
	\item $\Phi^{G \dagger}P\Phi$=$ - i\left( {{\phi ^T}{\tau _2}\phi  + {\phi ^\dag }\tau _2^\ast {\phi ^\ast }} \right)$=0;
	\item $\Phi^{G \dagger}\Gamma_i\Phi$=$- i\left( {{\phi ^T}{\tau _2}{\tau _i}\phi  - {\phi ^\dag }\tau _2^\ast \tau _i^ \ast {\phi ^ \ast }} \right)$, where ${\tau _2}{\tau _i} = {\left( {{\tau _2}{\tau _i}} \right)^T}$;
	\item $\Phi^{G \dagger}\Sigma_i\Phi$=$- i\left( {{\phi ^T}{\tau _2}{\tau _i}\phi  + {\phi ^\dag }\tau _2^\ast \tau _i^ \ast {\phi ^ \ast }} \right)$,
\end{enumerate}
where relations of
\begin{eqnarray}
{\left( {{\Phi ^{G\dag }}{\Gamma _i}\Phi } \right)^\dag } = {\Phi ^{G\dag }}{\Gamma _i}\Phi,
\quad
{\left( {{\Phi ^{G\dag }}{\Sigma _i}\Phi } \right)^\dag } = -{\Phi ^{G\dag }}{\Sigma _i}\Phi,
\end{eqnarray}
are satisfied. Therefore, interactions of scalars consist of $\Phi^\dagger\Phi$ as the scalar, $\Phi^\dagger \Gamma_i\Phi$ and $\Phi^{G \dagger}\Gamma_i\Phi$ as the vector and $\Phi^{G \dagger}\Sigma_i\Phi$ as the axial vector.  The mass and kinetic terms are described by the form of $\Phi^\dagger\Phi$.  Quartic couplings consist of $\left(\Phi^\dagger\Phi\right)^2$ and of the appropriate products out of $\Phi^\dagger\Gamma_i\Phi$, $\Phi^{G \dagger}\Gamma_i\Phi$ and $\Phi^{G \dagger}\Sigma_i\Phi$.  A Fierz identity relates these quartic terms. In fact, 
\begin{equation}
\left( {{\Phi ^\dag }{\Gamma _i}\Phi } \right)\left( {{\Phi ^\dag }{\Gamma _i}\Phi } \right) = \left( {{\Phi ^\dag }\Phi } \right)^2,
\end{equation}
is satisfied. Similarly, we find that 
\begin{eqnarray}
&&
\left( {{\Phi ^{G\dag }}{\Gamma _i}\Phi } \right)\left( {{\Phi ^\dag }{\Gamma _i}{\Phi ^G}} \right) = \left( {{\Phi ^{G\dag }}{\Sigma _i}\Phi } \right)\left( {{\Phi ^\dag }{\Sigma _i}{\Phi ^G}} \right)
={\left( {{\Phi ^\dag }\Phi } \right)^2},
\\
&&
\left( {{\Phi ^{G\dag }}{\Gamma _i}\Phi } \right)\left( {{\Phi ^\dag }{\Gamma _i}\Phi } \right)
=\left( {{\Phi ^{G\dag }}{\Sigma _i}\Phi } \right)\left( {{\Phi ^\dag }{\Gamma _i}\Phi } \right)
=\left( {{\Phi ^{G\dag }}{\Sigma _i}\Phi } \right)\left( {{\Phi ^\dag }{\Gamma _i}\Phi^G } \right)
=0.
\end{eqnarray}

As quartic couplings, it is sufficient to use $\left( {{\Phi ^\dag }\Phi } \right)^2$. Since $\Phi ^\dag \Phi = 2{\phi ^\dag }\phi$, the lagrangian of $\Phi$, $\mathcal{L}_\Phi$, is determined to be
\begin{eqnarray}
\mathcal{L}_\Phi
&=&
\frac{1}{2}\partial^\mu\Phi^\dagger\partial_\mu\Phi-V_\Phi,
\nonumber\\
V_\Phi
&=&
\frac{\mu^2_\Phi}{2}\Phi^\dagger\Phi+\frac{\lambda_\Phi}{4}\left(\Phi^\dagger\Phi\right)^2,
\end{eqnarray}
where $\mu_\Phi$ and $\lambda_\Phi$ stand for a mass and a quartic coupling, respectively. $\mathcal{L}_\Phi$ is also invariant under a phase transformation induced by $\phi^\prime=e^{-iq}\phi$ as a global $U(1)$ symmetry, where $q$ is real. In terms of $\Phi$, the invariance of $\mathcal{L}_\Phi$ is associated with the transformation of
\begin{equation}
{\Phi^\prime 
=
D_{ij}}\left( {\sigma ,\theta } \right)e^{-iqP}\Phi
=
e^{-i\sigma qP}D_{ij}\left( {\sigma ,\theta } \right)\Phi,
\end{equation}
where $P$ changes its sign for $\sigma=-1$ because of Eq.(\ref{Eq:GammaSigmaPRotation}), reflecting the fact that $P$ for $\phi^\ast$ is equal to $P$ for $\phi^\prime$.  This charge given by $P$ can be interpreted as a $\Phi$-number, which is referred to the \lq\lq dark number\rq\rq. It should be noted that the transformation of $\Phi^\prime = e^{-iq}\Phi$ such as in $U(1)$ of the standard model is incompatible with the definition of $\Phi$ because the transformed state $\Phi^\prime$ cannot satisfy the consistency condition of Eq.(\ref{Eq:ConsistencyCond}).

Although $\mathcal{L}_\Phi$ expressed in terms of $\Phi$ is nothing but a lagrangian expressed in terms of $\phi_{1,2}$, the $O(3)$ structure can be recognized if its complex conjugate is treated on the same footing. For example, the identity of $2\phi^\dag \phi=\sum\limits_{a = 1}^2(\phi^\dag_a \phi_a+\phi^T_a\phi^\ast_a)$ gives $\Phi^\dag\Phi$. In this simple case, it is invariant under the interchange of $\phi\leftrightarrow \phi^\ast$, which is caused by the parity operator $i{\tilde P}$ of Eq.(\ref{Eq:SigmaTildawithP}).

\subsection{Vectorlike fermion}\label{sec:spinor}
\subsubsection{Consistency with the Dirac fermion}\label{sec:Dirac_spinor}
For the Dirac fermion denoted by $\psi$, the fermionic $O(3)$ spinor, which is consistent with the Dirac spinor, should be composed of $\psi^C$ instead of $\psi^\ast$ to form $(\psi,\psi^C)^T$, where $\psi^C$ is a charge conjugation of $\psi$ defined by $\psi = C\bar \psi^T$.  The charge conjugation operator, $C$, satisfies that
\begin{equation}
{C^{ - 1}}{\gamma ^\mu }C =  - {\gamma ^{\mu T}},
\quad
{C^\dag } = {C^{ - 1}},
\quad
{C^T} =  - C,
\end{equation}
where $\gamma ^\mu$ ($\mu=0,1,2,3$) represents the Dirac gamma matrices.  The definition of the $O(3)$ spinor demands that $\psi^C\neq\psi$. Since the $O(3)$ spinor is required to be neutral, the possible candidate is a Majorana fermion, which is, however, excluded because of $\psi^C=\psi$. Our neutral $O(3)$ spinor carries the different $\boldsymbol{Z}_2$ charge; therefore, $\psi$ and $\psi^C$ are distinguished. 

Let us start with a pair of Dirac fields, $\psi_1$ and $\psi_2$, which form 
\begin{equation}
\xi=\left(
  \begin{array}{c}
     \psi_1   \\
     \psi_2  \\
  \end{array}
\right),
\end{equation}
as the $SO(3)$ doublet spinor and $\xi^C=(\psi^C_1,\psi^C_2)^T$  as its charge conjugate. In terms of $\xi$, the $O(3)$ doublet spinor is expressed to be $\Psi$:
\begin{equation}
\Psi = \left( \begin{array}{l}
\xi\\
\xi^C
\end{array} \right).
\end{equation}
The fermionic spinor behaves as the $O(3)$ spinor if $\Psi$ obeys the $O(3)$ transformation of $\Psi^\prime = D_{ij}(\sigma,\theta)\Psi$. The $O(3)$ spinor compatible with the transformation must satisfy Eq.(\ref{Eq:ConsistencyCondGeneral}).  For the fermionic $O(3)$ spinor, since $\Psi_{a+2}=\xi^C_a$=$C\gamma^{0T}\Psi^\ast_a$ holds, the parameter $\eta$ in Eq.(\ref{Eq:ConsistencyCondGeneral}) is given by
\begin{equation}
\eta=C\gamma^{0T},
\end{equation}
which turns out satisfy the necessary condition of $\eta^\ast\eta=I$.  As a result, we confirm that $\Psi$ is the $O(3)$ spinor, which is subject to the correct $O(3$) transformation: 
\begin{equation}
\Psi^\prime = D_{ij}\left(\sigma, \theta\right)\Psi.
\label{Eq:O3TransformationPsi}
\end{equation}
The use of $\psi^C_a$ instead of $\psi^\ast_a$ to form $\Psi$ assures the orthogonality of the $O(3)$ symmetry to the Lorentz symmetry.  It is obvious that the fermionic $O(3)$ spinor is vectorlike.  If it is not vectorlike, $\xi$ and $\xi^C$ have different chiralities so that $\xi^{C\prime}  = S^\ast_{ij}(-1,\theta)\xi$ given by Eq.(\ref{Eq:O3TransformationPsi}) cannot be satisfied.

\subsubsection{Lagrangian of the fermionic $O(3)$ spinor}\label{sec:lagrangian_fermionic_spinor}
Treating charge conjugates on the same footing, we use the familiar relations given by ${\overline {\psi_a}} i\cancel{\partial} \psi_a  = {\overline {\psi^C_a}} i\cancel{\partial} {\psi ^C_a}$ up to the total derivative and ${\overline {\psi_a}} \psi_a  =  {\overline {\psi^C_a}} \psi ^C_a$. The lagrangian of $\psi_a$ can be defined by
\begin{eqnarray}
\mathcal{L}_\Psi
&=&
\sum\limits_{a = 1}^2 {\overline {\psi_a}}\left( {i\cancel{\partial}  - m_\Psi} \right){\psi _a}  
\nonumber\\
&=&
\frac{1}{2}\sum\limits_{a = 1}^2 \left[ {\overline {\psi_a}}\left( {i\cancel{\partial}  - m_\Psi} \right){\psi _a} + {\overline {\psi _a^C}} \left( {i\cancel{\partial}  - m_\Psi} \right){\psi^C _a} \right]  
\nonumber\\
&=&
 \frac{1}{2}\bar \Psi \left( {i\cancel{\partial}  - m_\Psi} \right)\Psi,
\end{eqnarray}
where $m_\Psi$ is a mass of $\psi_a$. The invariance under the $O(3)$ transformation of $D_{ij}(\sigma.\theta)$ acting on $\Psi$ is satisfied by $\mathcal{L}_\Psi$ because of $\bar \Psi$=$\Psi^\dag\gamma^0$, where $\gamma^\mu$ commutes with $D_{ij}(\sigma,\theta)$. The possible bilinears including ${\bar \Psi}\Psi$ are expressed in terms of $\xi$ as follows:
\begin{eqnarray}
&&
\bar \Psi \Psi  = 2\bar \xi \xi,
\quad
\bar \Psi P\Psi  = 0,
\quad
\bar \Psi {\Gamma _i}\Psi  = 2\bar \xi {\tau _i}\xi,
\quad
\bar \Psi {\Sigma _i}\Psi  = 0,
\nonumber\\
&&
\bar \Psi {\gamma ^\mu }\Psi  = 0,
\quad
\bar \Psi {\gamma ^\mu }P\Psi  = 2\bar \xi {\gamma ^\mu }\xi,
\quad
\bar \Psi {\gamma ^\mu }{\Gamma _i}\Psi  = 0,
\quad
\bar \Psi {\gamma ^\mu }{\Sigma _i}\Psi  = 2\bar \xi {\gamma ^\mu }{\tau _i}\xi,
\end{eqnarray}
and those containing $\gamma_5$ are
\begin{eqnarray}
&&
\bar \Psi {\gamma _5}\Psi  = 2\bar \xi {\gamma _5}\xi,
\quad
\bar \Psi P{\gamma _5}\Psi  = 0,
\quad
\bar \Psi {\gamma _5}{\Gamma _i}\Psi  = 2\bar \xi {\gamma _5}{\tau _i}\xi,
\quad
\bar \Psi {\gamma _5}{\Sigma _i}\Psi  = 0,
\nonumber\\
&&
\bar \Psi {\gamma _5}{\gamma ^\mu }\Psi  = 2\bar \xi {\gamma _5}{\gamma ^\mu }\xi,
\quad
\bar \Psi {\gamma _5}{\gamma ^\mu }P\Psi  = 0,
\quad
\bar \Psi {\gamma _5}{\gamma ^\mu }{\Gamma _i}\Psi  = 2\bar \xi {\gamma _5}{\gamma ^\mu }{\tau _i}\xi,
\\
&&
\bar \Psi {\gamma _5}{\gamma ^\mu }{\Sigma _i}\Psi  = 0.
\nonumber
\end{eqnarray}

As for Yukawa coupling of $\Psi$, $\Phi$ accompanied by an $O(3)$-singlet Majorana fermion $N$ satisfying that $N^C=N$ can yield
\begin{equation}
\mathcal{L}_Y
=
-f\overline N \sum\limits_{a = 1}^2 {\phi _a^\dag {\psi _a}}  + \text{h.c.},
\end{equation}
where $f$ is a coupling constant. Considering relations of $\bar N\phi _a^\dag {\psi _a} = \overline {\psi _a^C} \phi _a^\ast N$ and $\overline {{\psi _a}} {\phi _a}N = \bar N\phi _a^T\psi _a^C$ to form $\Phi$ and $\Psi$,  $\mathcal{L}_Y$ turns out to be
\begin{equation}
\mathcal{L}_Y
=
-\frac{1}{2}
f{\bar N}\Phi ^\dag \Psi
+ \text{h.c.}.
\label{NPsiPhi}
\end{equation}
The invariance under the $O(3)$ symmetry is satisfied because $\Phi^\dag\Psi$ is a singlet of $O(3)$.  In $\mathcal{L}_Y$, $\Phi$ can be replaced by its G-conjugate $\Phi^G$.  If both couplings of $\Phi$ and $\Phi^G$ are present, the $U(1)$ symmetry is not respected.

The lagrangian for $N$ can be expressed in terms of $N_R$ and $\Psi_L$.  Since 
\begin{equation}
{\overline N} {\phi _a^\dag {\psi _a}}={\overline {N_R}} {\phi^\dag }{\xi_L}+{\overline {N_R}} {\phi^T}\xi_L^C+\text{h.c.},
\label{Eq:RightHandedN}
\end{equation}
the resulting lagrangian is
\begin{equation}
\mathcal{L}_Y
=
-f{\overline {N_R}} {\Phi ^\dag }\Psi_L
+ \text{h.c.}.
\label{Eq:VectorlikeFermion}
\end{equation}
In a practical sense, $N_R$ can be taken as a right-handed neutrino \cite{Kitabayashi2018} that couples to flavor neutrinos (to be denoted by $\nu$). The Majorana mass of $N_R$ generates very tiny masses of $\nu$ due to the seesaw mechanism \cite{Minkowski1977,Yanagida1979,Gell-Mann1979,Glashow1980,Mohapatra1980,Schechter1980}.

\subsection{Gauge boson}\label{sec:vector}
Let us start with a lagrangian for an $SO(3)$-triplet vector, $V_{i\mu}$, and an $SO(3)$-singlet vector $U_\mu$ accompanied by $\phi$ and $\xi$:
\begin{eqnarray}
\mathcal{L}
&=&
- \frac{1}{4}V_i^{\mu \nu }{V_{i\mu \nu }}
- \frac{1}{4}U^{\mu \nu }{U_{\mu \nu }}
+
\mathcal{L}_\phi
+
\mathcal{L}_\xi,
\nonumber\\
\mathcal{L}_\phi
&=&
{\left| {\left( {{\partial _\mu } + ig{V_\mu } + ig^\prime\frac{Y_U}{2}{U_\mu }} \right)\phi } \right|^2},
\label{Eq:GaugeLag}
\\
\mathcal{L}_\xi
&=&
{\bar \xi {\gamma ^\mu }\left( {i{\partial _\mu } - g{V_\mu } - g^\prime\frac{Y_U}{2}{U_\mu }} \right)\xi },
\nonumber
\end{eqnarray}
for $g$ and $g^\prime$ as coupling constants and $Y_U$ stands for a $U(1)$ charge, where mass terms are omitted for simplicity and
\begin{equation}
{V_{i\mu \nu }}
=
{\partial _\mu }{V_{i\nu }} - {\partial _\nu }{V_{i\mu }} + g{\varepsilon _{ijk}}{V_{j\mu }}{V_{k\nu }},
\quad
{U_{\mu \nu }}
=
{\partial _\mu }{U_{\nu }} - {\partial _\nu }{U_{\mu }},
\quad
{V_\mu }
=
\frac{{{\tau _i}}}{2}{V_{i\mu }}.
\end{equation}
To introduce $\Phi$ and $\Psi$, the relations to the complex conjugate of $\mathcal{L}_{\phi,\xi}$:
\begin{eqnarray}
&&
 {\left| {\left( {{\partial _\mu } - igV_\mu ^ \ast   - ig^\prime\frac{Y_U}{2}{U_\mu }} \right){\phi ^\ast}} \right|^2}
=
{\left| {\left( {{\partial _\mu } + ig{V_\mu } + ig^\prime\frac{Y_U}{2}{U_\mu }} \right)\phi } \right|^2},
\nonumber\\
&&
\overline {{\xi ^C}} {\gamma ^\mu }\left( {i{\partial _\mu } + gV_\mu ^T + g^\prime\frac{Y_U}{2}{U_\mu }} \right){\xi ^C}
=
\bar \xi {\gamma ^\mu }\left( {i{\partial _\mu } - g{V_\mu } - g^\prime\frac{Y_U}{2}{U_\mu }} \right)\xi,  
\end{eqnarray}
are explicitly used. 

Now, after $V_\mu ^T = V_\mu ^\ast $ is used for $\xi^C$, $\tau_i$ for $\phi$ and $\xi$ can be replaced by $\text{diag.}(\tau_i, -\tau^\ast_i)$ for $\Phi$ and $\Psi$, which is nothing but $\Sigma_i$. As a result, we find that
\begin{eqnarray}
\mathcal{L}_\phi
&=&
\frac{1}{2}{\left| {\left( {{\partial _\mu } + ig{\boldsymbol{V}_\mu } + ig^\prime\frac{Y_U}{2}{\boldsymbol{U}_\mu }} \right) \Phi } \right|^2},
\nonumber\\
\mathcal{L}_\xi
&=&
\frac{1}{2}{\bar \Psi {\gamma ^\mu }\left( {i{\partial _\mu } - g{\boldsymbol{V}_\mu } - g^\prime\frac{Y_U}{2}{\boldsymbol{U}_\mu }} \right)\Psi },
\end{eqnarray}
where 
\begin{equation}
\boldsymbol{V}_\mu= \frac{\Sigma _i}{2}V_{i\mu },
\quad
\boldsymbol{U}_\mu= PU_\mu.
\end{equation}
Hereafter, $\Psi$ is assumed to possess the same $Y_U$ as $\Phi$ so as to yield the vanishing $U(1)$ charge of $\Phi^\dag\Psi$ in $\mathcal{L}_Y$.  In $\mathcal{L}_{\phi,\xi}$, (${\Phi ^\dag }{\Sigma _i}{\partial _\mu }\Phi$,  ${\bar \Psi}\gamma^\mu\Sigma_i\Psi$) and (${\Phi ^\dag }P{\partial _\mu }\Phi$, ${\bar \Psi}\gamma^\mu P\Psi$), respectively, transform as the axial vector and the pseudoscalar as indicated by Eqs.(\ref{Eq:Rotation}) and (\ref{Eq:GammaSigmaPRotation}). It is concluded that
\begin{itemize}
	\item $V_{i\mu}$ is the axial vector transformed by $V^\prime_{k\mu}$ = $\sigma T_{ij}(\sigma,\theta)_{k\ell}V_{\ell\mu}$;
	\item $U_\mu$ is the pseudoscalar transformed by $U^\prime_\mu$ = $\sigma U_\mu$,
\end{itemize}
to get the $O(3)$-invariant lagrangian.  This feature may represent a remnant of the finite $O(3)$ transformation.  It is understood that the gauged $SU(2)\times U(1)$ symmetry is generated by $\Sigma_i$ and $P$ linked to $SO(3) \times \boldsymbol{Z}_2$.

Generally speaking, $V_{i\mu}$ should transform as the axial vector of $O(3)$. If $V_{i\mu}$ transforms as the vector of $O(3)$, the nonabelian term of ${\varepsilon _{ijk}}{V_{j\mu }}{V_{k\nu }}$ transforms as the axial vector of $O(3)$ because this term exhibits an antisymmetric configuration as the cross product of two vectors. The nonabelian term, which is the axial vector, is inconsistent with the abelian term of ${\partial _\mu }{V_{i\nu }} - {\partial _\nu }{V_{i\mu }}$, which is the vector.  The consistent assignment to $V_{i\mu }$ is clearly to use the axial vector of $O(3)$.  There is no such a constraint on matter field in the triplet representation because the similar nonabelian term involved in the covariant derivative is always consistent with the abelian term.  For $U_\mu$ as the pseudoscalar of $O(3)$, the consistency is related to the fact that the eigenvalues of the $U(1)$ charge of the $O(3)$-transformed spinor are opposite in signs to the original spinor as stressed in Eq.(\ref{Eq:PseudoP}).

\section{Dark Gauge Symmetry}\label{sec:darksector}

In this section, we argue how our $O(3)$ spinors together with the gauge bosons are suitable to describe dark sector of particle physics. Our $O(3)$ spinors are restricted to be neutral and this restriction may be good news for dark matter, which is considered to be neutral. We do not intend to show detailed numerical analyses of effects of our $O(3)$ spinors on dark matter but how the spinorial $O(3)$ structure affects the dark sector physics, especially, based on a new realization of the $O(3)$ gauge symmetry. The $O(3)$ symmetry can be elevated to the $SU(2)\times U(1)$ gauge symmetry with $V_{i\mu}$ in the axial vector representation and $U_\mu $ in the pseudoscalar representation, which couple to neutral $O(3)$ spinors. The $U(1)$ symmetry is related to the $\boldsymbol{Z}_2$ parity of $O(3)$. Since our dark fermions are vectorlike, no chiral anomalies are present. Our gauge symmetry may serve as a dark gauge symmetry \cite{Arkani-Hamed2009,Zhang2010}, which includes a dark photon as a dark gauge boson \cite{Dobrescu2005,Pospelov2008}.  Our gauged $O(3)$ model provides a vector portal dark matter \cite{Hambye2009}. 

The dark photon sometimes remains massless \cite{Dobrescu2005}.  As the worst case, $\Phi$ and $\Psi$, provided that they are dark charged, are annihilated into the massless dark photon if the dominance of dark matter over dark antimatter is not effective. In the rest of discussions, we assume that this dominance is based on the universal mechanism over the universe \cite{Kaplan1992,Dodelson1992} so that dark matter remains in the dark sector after the annihilation is completed. For instance, since our interactions may contain $N_R$ as a right-handed neutrino in $\mathcal{L}_Y$, the dominance of matter over antimatter can be based on the leptogenesis utilizing the lepton number violation due to $N_R$ \cite{Fukugita1986}, which in turn produces a dark matter particle-antiparticle asymmetry in the dark sector \cite{Cosme2005,Kaplan2009}. Reviews on the dark matter and baryon abundances are presented in Ref.\cite{Petraki2013,Zurek2014}, where further references can be found.

Dark matter scenarios based on the gauged $O(3)$ symmetry as a maximal gauge symmetry provide the similar scenarios to the existing ones \cite{Alexander2016}. However, the essential difference from the existing scenarios lies in the fact that our dark matter and dark gauge bosons are theoretically forced to be all neutral once the spinorial structure of $O(3)$ is built in the dark sector.  From this feature, the kinetic mixing with the photon \cite{Holdom1986} is not a primary mixing with the standard model gauge bosons.  Instead, the mixing with the $Z$ boson becomes important and arises as an induced loop effect that allows $Z$ to couple to intermediate flavor neutrinos.  Another difference is that our gauge bosons do not transform as the triplet vector for $V_{i\mu}$ and the singlet scalar for $U_\mu$ but as the axial vector and the pseudoscalar, respectively. 

\begin{figure*}[!htbp]
\begin{center}
\includegraphics[width=15cm]{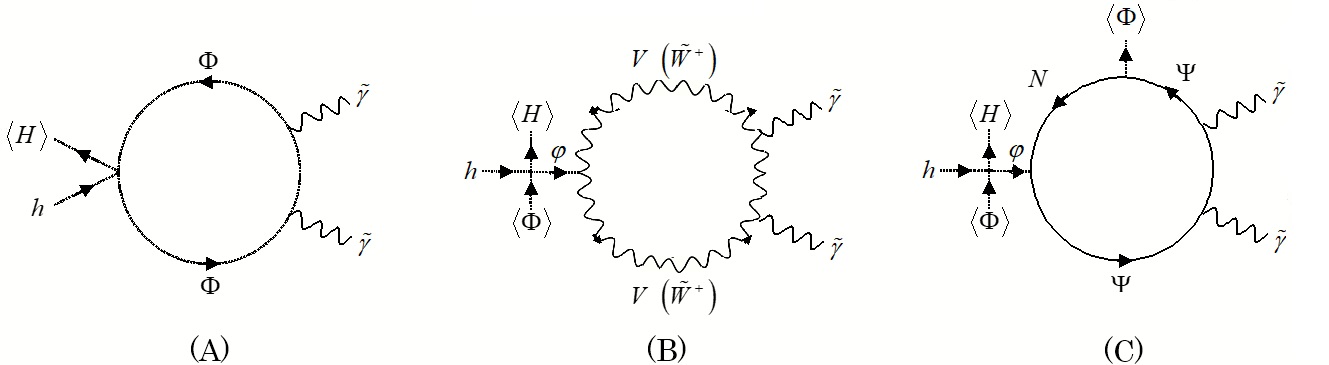}
\end{center}
\caption{The decay of $h\to\tilde \gamma\tilde \gamma$ in (A) unbroken $U(1)$ symmetry, (B) spontaneously broken $O(3)/U(1)$ and $O(3)$ symmetries, respectively, for the $V$-loop and the $\tilde W^+$-loop, and (C) spontaneously broken $O(3)/U(1)$ symmetry.}
\label{Fig:h-2xgamma_tilde}
\end{figure*}

\subsection{Unbroken $U(1)$ symmetry}\label{sec:U1darkphoton}
The simplest realization of the dark photon denoted by ${\tilde \gamma}$ can be based on the $U(1)$ gauge model. When the $U(1)$ symmetry is not spontaneously broken, the dark photon stays massless. The dark photon interacts with the standard model particles via $H$. The relevant part of the lagrangian of $\Phi$ including $H$ is
\begin{eqnarray}
\mathcal{L}
&=&
\frac{1}{2}{\bar \Psi \left[{\gamma ^\mu }\left( {i{\partial _\mu } - g^\prime\frac{Y_\Phi}{2}{\boldsymbol{U}_\mu }} \right)-m_\Psi\right]\Psi }
+
\mathcal{L}_Y
+
\frac{1}{2}{\left| {\left( {{\partial _\mu }  + ig^\prime\frac{Y_\Phi}{2}{\boldsymbol{U}_\mu }} \right) \Phi } \right|^2}
-
V_S,
\nonumber\\
{V_S} 
&=&
{\mu ^2_H}{H^\dag }H
+
\lambda_H {\left( {{H^\dag }H} \right)^2}
+
\frac{{\mu _\Phi ^2}}{2}{\Phi ^\dag }\Phi 
+
\frac{{{\lambda _\Phi }}}{4}{\left( {{\Phi ^\dag }\Phi } \right)^2}
+
\frac{{{\lambda _{H\Phi }}}}{2}{H^\dag }H{\Phi ^\dag }\Phi,
\label{Eq:VScalars}
\end{eqnarray}
where $\mu_{H,\Phi}$ and $\lambda_{H,\Phi}$, respectively, stand for mass parameters and quartic couplings. The physical Higgs scalar $h$ is parameterized in $H$ = $(0,(v+h)/\sqrt{2})$, where $v$ is a VEV of $H$.

Since $\Phi$ couples to the $U(1)$ gauge boson but does not couple to the standard model gauge bosons, the decay of $h$ proceeds via 
\begin{itemize}
	\item $h\to {\tilde \gamma}{\tilde \gamma}$ triggered by the one loop of $\Phi$,
\end{itemize}
which is generated by $\Phi^\dag\Phi H^\dag H$ \cite{Biswas2015}. The decay of $h\to\tilde \gamma\tilde \gamma$ is depicted in FIG.\ref{Fig:h-2xgamma_tilde} including other two cases to be discussed.  The dark matter consists of the lightest $O(3)$ spinors. Because of $\Phi\to\Psi{\bar \nu}$ or $\Psi\to\Phi\nu$ induced by $\mathcal{L}_Y$, depending upon the magnitude relation of masses of $\Phi$ and $\Psi$, the lightest $O(3)$ spinor is either $\Phi$ or $\Psi$, which has two degenerated components. 

\subsection{Spontaneously broken $O(3)/U(1)$ symmetry}\label{sec:SO3darkphoton}
The next example is based on the $O(3)$ gauge symmetry without the $U(1)$ symmetry, which we refer as an {\lq\lq$O(3)/U(1)$\rq\rq} symmetry. The relevant part of the lagrangian of $\Phi$ and $\Psi$ is
\begin{eqnarray}
\mathcal{L}
&=&
\frac{1}{2}{\bar \Psi \left[{\gamma ^\mu }\left( {i{\partial _\mu } - g{\boldsymbol{V}_\mu }} \right)-m_\Psi\right]\Psi }
+
\mathcal{L}_Y
+
\frac{1}{2}{\left| {\left( {{\partial _\mu }  + ig{\boldsymbol{V}_\mu }} \right) \Phi } \right|^2}
-
V_S.
\label{Eq:LforSO3}
\end{eqnarray}
In addition to $\mathcal{L}_Y$, another $\mathcal{L}_Y$ as $\mathcal{L}^\prime_Y$ given by 
\begin{equation}
\mathcal{L}^\prime_Y
=
-f^\prime{\overline {N_R}} {\Phi ^{G\dag} }\Psi_L
+ \text{h.c.},
\label{Eq:VectorlikeFermionG}
\end{equation}
can be included because of the absence of $U(1)$.\footnote{Needless to say, the condition of $\psi\neq \psi^C$, which is ascribed to $U(1)$, is also fulfilled by the $SO(3)$ spinor property that ensures $\psi\neq\psi^C$} To generate spontaneous breakdown, we assume that $\phi_2$ in $\phi$ acquires a VEV.\footnote{For $\langle\Phi\rangle\propto \Gamma_3$ leading to $U(1)$ as a residual symmetry, see Ref.\cite{Khoze2014}.} For $\mu^2_\Phi<0$ in Eq.(\ref{Eq:LforSO3}), $\phi_2$ develops a VEV that spontaneously breaks $O(3)/U(1)$.  The physical scalar denoted by $\varphi$ is parameterized in $\phi$ = $(0,(v_\Phi+\varphi)/\sqrt{2})$ for $v_\Phi$ as a VEV of $\Phi$.

Three of the four real components in $\Phi$ are absorbed into three dark gauge bosons, which become massive. The remaining scalar is a massive $\varphi$.  Interactions of scalars provide the following results:
\begin{itemize}
	\item $h$ and $\varphi$ are mixed via the mass term of $M_{mix}$ calculated to be:
\begin{equation}
{M_{mix}} 
= 
\left( {\begin{array}{*{20}{c}}
  {m_h^2}&{{\lambda _{\Phi H}}{v_\Phi }v} \\ 
  {{\lambda _{\Phi H}}{v_\Phi }v}&{m_\varphi ^2} 
\end{array}} \right),
\label{Eq:HiggsMixing}
\end{equation}
where ${m^2_h} = 2{\lambda_H}v^2$ and ${m^2_\varphi} = 2{\lambda _\Phi}v^2_\Phi$;
	\item The massive dark gauge bosons have a degenerated mass of $g v_\Phi/2$ and play a r\^{o}le of $\tilde \gamma$;
	\item $h\to\tilde \gamma\tilde \gamma$ via the one-loop by $V$ (FIG.\ref{Fig:h-2xgamma_tilde}-(B) and by $\Psi$ and $N_R$ (FIG.\ref{Fig:h-2xgamma_tilde}-(C)) if the decay is kinematically allowed.
\end{itemize}
The mixing of $h$ and $\varphi$ should be carefully examined to be consistent with the currently established Higgs phenomenology \cite{Ahlers2008,Weinberg2013,Clarke2014,Garcia2014}. For a reference value of the Higgs mixing angle $\theta_H$ associated with Eq.(\ref{Eq:HiggsMixing}), we quote that $\left| \tan\theta_H\right|\lesssim 2.2\times 10^{-3}(v_\Phi/10~{\text {GeV}})$ derived from the constraint on invisible Higgs decays.  The dark matter consists of $\varphi$ and $\Psi$ because $\tilde \gamma$ may dissolve into $\nu{\bar \nu}$ (see Sec.\ref{sec:darkgaugebosons}).  The VEV of $\langle H\rangle$ induces a coupling of ${\bar \nu}\varphi^\dag\psi_{1,2}$, which yields 
\begin{itemize}
	\item $\varphi\to\psi_{1,2}{\bar \nu}$/$\psi_{1,2}\nu$ or $\psi_{1,2}\to\varphi\nu$/$\varphi{\bar \nu}$.
\end{itemize}
The lightest $O(3)$ spinor is either $\varphi$ or $\psi_{1,2}$.

\subsection{Spontaneously broken $O(3)$ symmetry}\label{sec:O3darkphoton}
The relevant part of the lagrangian of $\Phi$ and $\Psi$ is
\begin{eqnarray}
\mathcal{L}
&=&
\frac{1}{2}{\bar \Psi \left[{\gamma ^\mu }\left( {i{\partial _\mu } - g{\boldsymbol{V}_\mu }  - g^\prime\frac{Y_\Phi}{2}{\boldsymbol{U}_\mu }} \right)-m_\Psi\right]\Psi }
+
\mathcal{L}_Y
\nonumber\\
&+&
\frac{1}{2}{\left| {\left( {{\partial _\mu }  + ig{\boldsymbol{V}_\mu } - g^\prime\frac{Y_\Phi}{2}{\boldsymbol{U}_\mu }} \right) \Phi } \right|^2}
-
V_S.
\label{Eq:LforO3}
\end{eqnarray}
In the case that the $O(3)$ symmetry is spontaneously broken, if the appropriately defined $U(1)$ charge is not carried by $\phi_2$, there remains a massless gauge boson chosen to be $\tilde \gamma$ as a mixed state of $V_{3\mu}$ and $U_\mu$ as in the standard model \cite{Davoudiasl2013}. Since the $U(1)$ invariance is assumed in the $O(3)$ symmetry, $\mathcal{L}^\prime_Y$ cannot be included. To be specific, without the loss of the generality, $Y_U=1$ for $\Phi$ can be chosen. Since everything is the same as in the standard model, we use the tilde to denote dark gauge bosons with the obvious notations.

The massless dark photon identified with $\tilde A$ can couple to dark charged particles.  For $\Psi$, $\psi_1$ is a dark charged particle while $\psi_2$ is a dark neutral particle because $Y_\Psi=Y_\Phi$ is taken in Eq.(\ref{Eq:LforO3}). The remaining dark charged particles are $\tilde W^\pm$. The decay of $h\to {\tilde \gamma}{\tilde \gamma}$ needs both $\varphi$ and $\tilde \gamma$ to couple to intermediate particles. The possible intermediate particles are $\tilde W^\pm$ since $\varphi$ does not couple to the dark charged $\psi_1$ while $\tilde \gamma$ does not couple to the dark neutral $\phi_2$ and $\varphi$. The diagram as in FIG.\ref{Fig:h-2xgamma_tilde}-(C) is absent.  The decay of $h$ proceeds via $\tilde W^\pm$: 
\begin{itemize}
	\item $h\to {\tilde \gamma}{\tilde \gamma}$ via the one loop by ${\tilde W}^\pm$,
\end{itemize}
because ${\tilde W}^\pm$ interact with both $\varphi$ and $\tilde \gamma$ as shown in FIG.\ref{Fig:h-2xgamma_tilde}-(B). 

Our dark matter consists of $\varphi$ and $\psi_{1,2}$. For other candidates of $\tilde W^\pm$ and $\tilde Z$, $\tilde W^\pm$ disappear because $\tilde W^+\tilde W^-$ are annihilated into $\tilde \gamma \tilde \gamma$ and $\tilde Z$ decays into $\nu{\bar \nu}$ owing to the $\tilde Z$ mixing with the $Z$ boson (see Sec.\ref{sec:darkgaugebosons}). There arises an interaction between $\varphi$ and $\psi_2$, which is ${\bar \nu}\psi_2\varphi$ induced by the collaboration of ${\overline {N_R}}\varphi\psi_2$ and ${\bar \nu}hN_R$.  Due to this interaction, the heavier $O(3)$ spinor will decay according to the decay mode of either $\varphi\to\psi_2{\bar \nu}$/$\psi_2\nu$ or $\psi_2\to\varphi\nu$/$\phi_2{\bar \nu}$. For $\psi_1$, these decay modes yielding $\nu$ are absent because $\mathcal{L}^\prime_Y$ is absent. Since $\psi_{1,2}$ are degenerated, $\psi_1$ is stable. The dark charged ${\tilde W}^\pm$ are annihilated into dark photons while $\tilde Z$ may decay into $\ell^+\ell^-$, $\nu{\bar \nu}$ and so on. We conclude that 
\begin{itemize}
	\item our dark matter consists of $\psi_1$ and the lighter spinor, which is either $\varphi$ or $\psi_2$;
	\item $\tilde \gamma$ has a direct coupling to the dark charged $\psi_1$ while $\varphi$ has a direct coupling to the dark neutral $\psi_2$.
\end{itemize}
The recent discussions on the phenomenology of the massless dark photon of this type can be found in Ref.\cite{Altmannshofer2015,Dunsky2019}. 

\subsection{Dark gauge bosons}\label{sec:darkgaugebosons}
The dark gauge bosons imported into these three subsections consist of
\begin{itemize}
	\item the massless $U$ boson identified with $\tilde \gamma$ in Sec.\ref{sec:U1darkphoton};
	\item three degenerated massive $V$ bosons identified with $\tilde \gamma$ in Sec.\ref{sec:SO3darkphoton};
	\item the massless ${\tilde A}$ identified with $\tilde \gamma$ and three massive bosons of ${\tilde W^\pm}$ and ${\tilde Z}$ in Sec.\ref{sec:O3darkphoton}.
\end{itemize}
It is a general feature that $\tilde \gamma$ cannot possess the hypercharge associated with the $U(1)_Y$ symmetry of the standard model, which enables $\tilde \gamma$ to kinematically mix with the photon. Instead, there appears the mixing between $\tilde \gamma$ and $Z$. 

\begin{figure*}[b]
\begin{center}
\includegraphics[width=15cm]{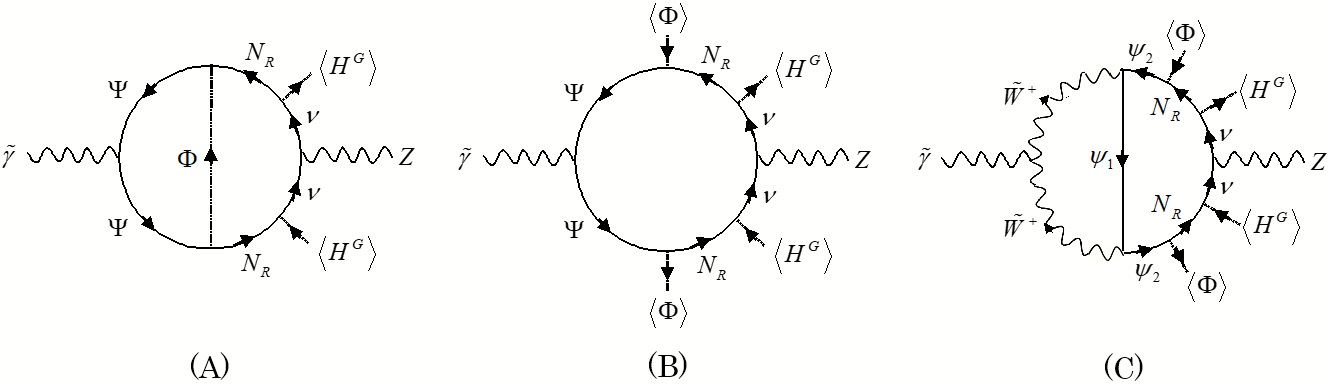}
\end{center}
\caption{The $\tilde \gamma$-$Z$ mixing in (A) unbroken $U(1)$ symmetry, where $\Phi$ and $\Psi$ can be interchanged, (B) spontaneously broken $O(3)/U(1)$ symmetry and (C) spontaneously broken $O(3)$ symmetry, where $\tilde W^+$ and $\psi_1$ can be interchanged.}
\label{Fig:gamma_tilde-Z}
\end{figure*}

\begin{table*}[!htbp]
\caption{\label{table1}Particle content of the dark sector.}
\begin{tabular}{|c||c|c|c|}
\hline
model	 & Sec.\ref{sec:U1darkphoton} & Sec.\ref{sec:SO3darkphoton} & Sec.\ref{sec:O3darkphoton} \\ \Hline
gauge symmetry & $U(1)$ & $O(3)/U(1)$ & $O(3)$\\ \cline{2-4}
 & unbroken & \multicolumn{2}{c|}{spontaneously broken}\\ \hline 
dark matter candidates & $\psi_{1,2}$ or $\phi_{1,2}$ & $\psi_{1,2}$ or $\varphi$ & $\psi_1$/$\psi_2$ or $\varphi$\\ \hline
dark charge & charged $\psi_{1,2},\phi_{1,2}$& --- & charged $\psi_1$, neutral $\psi_2,\varphi$ \\ \hline
dark photon $\tilde \gamma$ & massless $U$ & massive $V^{1,2,3}$ (degenerated) & massless $\tilde A$ \\ \hline
\end{tabular}
\end{table*}

The mixing of $\tilde \gamma$ with $Z$ arises as the following loop effects due to
\begin{itemize}
	\item two loops by $\Psi$, $N_R$ and $\nu$ in the external loop involving the internal exchange of $\Phi$ for Sec.\ref{sec:U1darkphoton} or one loop with the same external loop as in Sec.\ref{sec:U1darkphoton} but involving the nonvanishing $\langle\Phi\rangle$ for Sec.\ref{sec:SO3darkphoton};
	\item two loops by $\tilde W^+$, $\psi_2$, $N_R$ and $\nu$ in the external loop involving the internal exchange of $\psi_1$ for Sec.\ref{sec:O3darkphoton}, where $\tilde \gamma$ only couples to $\psi_1$, which cannot couple to $N_R$ because of the absence of $\mathcal{L}^\prime_Y$.
\end{itemize}
These induced couplings are depicted in FIG.\ref{Fig:gamma_tilde-Z}.\footnote{The Furry theorem is applied to quantum corrections containing the dark photons, whose contributions can be evaluated according to the ordinary rules since both particle and antiparticle are involved in the $O(3)$ spinors.}  The similar mixing with $Z$ is applied to ${\tilde Z}$.  Due to the same one-loop coupling as the $\tilde \gamma$-$Z$ mixing but one of $\langle H^G\rangle$'s is replaced with $h$, the $h$ decay proceeds via
\begin{itemize}
	\item $h\to Z\tilde \gamma$, which is necessarily induced for the massless $\tilde \gamma$.
\end{itemize}
The particle content of each realization of the dark gauge symmetry is summarized in TABLE.\ref{table1}.

\begin{figure}[t]
\begin{center}
\includegraphics[width=7.7cm]{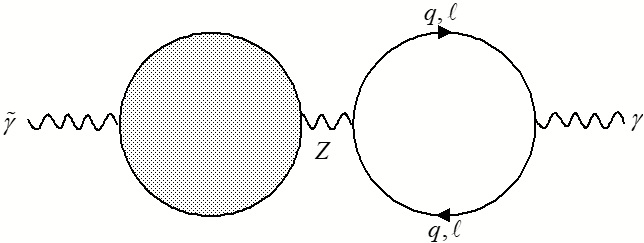}
\end{center}
\caption{The $\tilde \gamma$-$\gamma$ mixing, where the $\tilde \gamma$-Z mixing (shaded circle) is connected by $Z$ to the loop by quarks ($q$) and leptons ($\ell$).}
\label{Fig:gamma_tilde-gamma}
\end{figure}

For the massless $\tilde \gamma$, the mixing between $\tilde \gamma$ and $Z$ is restricted to the kinetic mixing.  For the massive $\tilde \gamma$, the induced decay of $\tilde \gamma$: $\tilde \gamma\to \nu{\bar \nu}$ is possible to occur via the intermediate $Z$ boson decaying into $\nu\bar\nu$. Considering higher loop effects involving the induced $\tilde \gamma$-$Z$ mixing, we have the kinetic $\tilde \gamma$-$\gamma$ mixing, which is generated by the $Z$-boson exchanged between the loop(s) giving the $\tilde \gamma$-$Z$ mixing and the one loop by quarks and leptons, which finally couples to the photon (FIG.\ref{Fig:gamma_tilde-gamma}).  Our kinetic $\tilde \gamma$-$\gamma$ mixing turns out to be further suppressed by the $Z$ boson mass.

\section{Scalar dark matter}\label{sec:phenomenology}
In this section, we would like to show results of phenomenological analyses on our dark matter, which include the estimation of relic abundance and nucleon spin-independent cross section of dark matter.  Our dark matter can be the $SO(3)$-doublet scalar $\phi$ in $\Phi$ and spinor $\psi$ in $\Psi$.  To see the feasibility of three $O(3)$ scenarios discussed in Sec.\ref{sec:darksector}, we simply choose $\phi$ as the candidate of the dark matter, which is lighter than $\psi$. In the present discussion, we concentrate on dark matter phenomenology based on the $O(3)$ model characterized by the $U(1)$ gauge symmetry presented in Sec.\ref{sec:U1darkphoton}, whose gauge boson plays a r\^{o}le of the dark photon. The model includes the dark scalar $\phi$, the dark Dirac spinor $\psi$ and the dark photon $\tilde \gamma$ as well as the Higgs scalar $H$ as can be seen from Eq.(\ref{Eq:VScalars}). 

Before proceeding to the phenomenological analyses, we discuss constraints on the dark matter, which mainly arise from the consistency with the dark matter contribution to the Higgs invisible decay.  Another known constraint comes from the consistency with the electroweak precision parameters, which are hardly modified since our dark matter being the $O(3)$ spinors is forbidden to directly couple to any of the standard model particles (SM).  It should be noted that the existing inert weak $SU(2)$ doublet, apparently similar to our $O(3)$ doublet, inevitably modifies electroweak precision parameters so that appropriate constraints are imposed on the inert weak $SU(2)$ doublet \cite{Bhattacharya2019}. 

\subsection{Invisible Higgs decay}\label{sec:invisible}
The direct decay of $H$ into dark matter is induced by the coupling of $\Phi^\dag\Phi H^\dag H$, where the mass of $\Phi$ is given by $m^2_\Phi = \mu _\phi ^2 + \lambda_{H\Phi}{v^2}/2$ and the decay width is estimated to be:
\begin{equation}
\Gamma\left(h\to\phi^\ast\phi\right)=\frac{\lambda^2_{H\phi} v^2}{16\pi m_h}\left(1-\frac{4m^2_\Phi}{m^2_h}\right)^{1/2},
\end{equation}
where $\phi$ runs over two degenerated scalars of $\phi_{1,2}$, provided that $m_\Phi < m_h/2$.  For $m_\Phi \gtrsim m_h/2$, the one-loop induced decay of $h\to{\tilde \gamma}{\tilde \gamma}$ by the $\Phi$-loop as in FIG.\ref{Fig:h-2xgamma_tilde}-(A) becomes a dominant decay mode. Considering the effect from both $\phi_1$ and $\phi_2$, we obtain its decay width calculated to be: 
\begin{equation}
\Gamma\left(h\to\tilde\gamma\tilde\gamma\right)=\left|\frac{\alpha^\prime\lambda_{H\phi} v}{8\pi^{3/2}}\frac{m^{3/2}_h}{m^2_\Phi}F\left(\frac{m^2_h}{4m^2_\Phi}\right)\right|^2,
\end{equation}
where $\alpha^\prime = g^{\prime 2}/4\pi$ and 
\begin{equation}
F\left(x\right)=-\left[x-f\left(x\right)\right]x^{-2},
\end{equation}
with
\begin{equation}
f\left( x \right) = \left\{ \begin{gathered}
  {\arcsin ^2}\left( {\sqrt x } \right){\text{ for }}x \leqslant 1 \hfill \\
   - \frac{1}{4}{\left( {\log \frac{{1 + \sqrt {1 - {x^{ - 1}}} }}{{1 - \sqrt {1 - {x^{ - 1}}} }} - i\pi } \right)^2}{\text{ for }}x > 1 \hfill \\ 
\end{gathered}  \right..
\end{equation}
The invisible Higgs decay satisfies that $\Gamma\left(h\to \textrm{invisible}\right)$ = $\Gamma\left(h\to\phi^\ast\phi\right)+\Gamma\left(h\to\tilde\gamma\tilde\gamma\right)$.  The total Higgs width ($\Gamma^{TOT}$) turns out to be $\Gamma^{TOT} = \Gamma _{STD} + \Gamma\left(h\to \textrm{invisible}\right)$, where $\Gamma _{STD}$ stands for the theoretically expected value of the Higgs total width estimated to be about 4.2 MeV \cite{Heinemeyer2013,CMSHIGGS2018}. 

\begin{figure}[t]
\begin{center}
\includegraphics{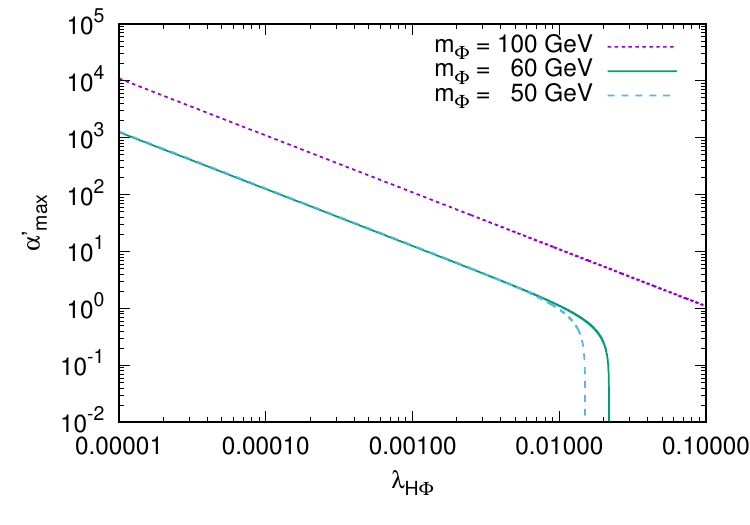} 
\includegraphics{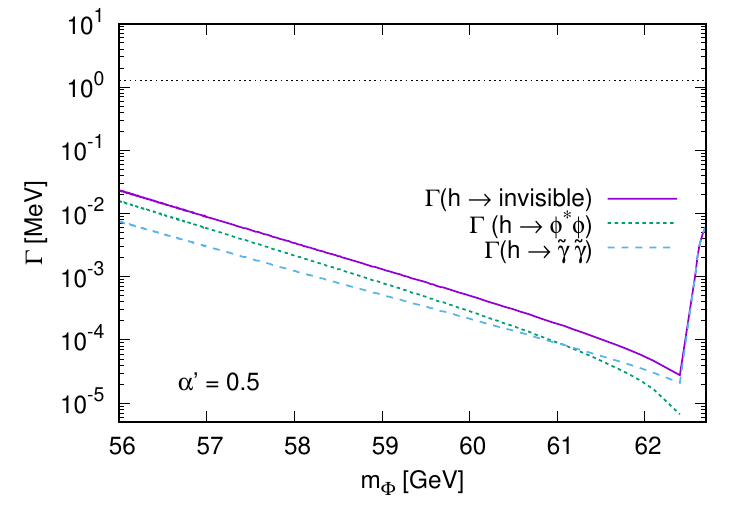} 
\caption{Allowed maximal values of $\alpha^\prime$, $\alpha^\prime_\textrm{max}$, to satisfy Eq.(\ref{Eq:HiggsInvisible}) in the left panel and prediction of $\Gamma\left( h\to \textrm{invisible}\right)$ in the right panel at $\alpha^\prime = 0.5$, where the dotted horizontal lines show $\Gamma \left( {h \to {\text{invisible}}} \right) = 1.3$ MeV from Eq.(\ref{Eq:HiggsInvisible}) for the allowed combinations of $\lambda_{H\Phi}$ and $m_\Phi$ to satisfy $\Omega h^2 = 0.120 \pm 0.001$ and $\sigma_{SI} \le 4.1 \times 10^{-47}$ cm$^2$.}
\label{fig:fig_alpha_dash}
\end{center}
\end{figure}

The branching ratio $\textrm{Br}\left(h\to \textrm{invisible}\right)$, which is experimentally constrained as $\textrm{Br}\left(h\to \textrm{invisible}\right) < 24\%(\equiv B_{MAX})$ \cite{RPP2018}, is translated into
\begin{equation}
{\Gamma \left( {h \to {\text{invisible}}} \right) < \frac{{{B_{MAX}}}}{{1 - {B_{MAX}}}}{\Gamma _{STD}}},
\end{equation}
leading to
\begin{equation}
\Gamma \left( {h \to {\text{invisible}}} \right) < 1.3~\textrm{MeV}.
\label{Eq:HiggsInvisible}
\end{equation}
The parameters, $\lambda_{H\Phi}$ and $\alpha^\prime$, are so constrained to satisfy Eq.(\ref{Eq:HiggsInvisible}) for a given dark matter mass.  To see the restriction on $\alpha^\prime$, we depict allowed maximal values of $\alpha^\prime$, $\alpha^\prime_\textrm{max}$, in the left panel of FIG.\ref{fig:fig_alpha_dash}, where $m_\Phi=50,60,100$ GeV and $0.00001 \le \lambda_{H\Phi} \le 0.1$ while the right panel shows $\Gamma\left( h\to \phi^\ast\phi,\tilde\gamma\tilde\gamma \right)$ as well as their sum giving $\Gamma \left( {h \to {\text{invisible}}} \right)$ at $\alpha^\prime = 0.5$ for $56.0 \lesssim m_\Phi [\textrm{GeV}] \lesssim 62.7$, which is the suggested range of $m_\Phi$ found in the next subsection to reproduce the observed value of $\Omega h^2$ and the experimental constraint on $\sigma_{SI}$ (see FIG.\ref{fig:lamHphi_m_with_fraction}). We find that $\Gamma\left(h\to \textrm{invisible}\right)\lesssim 10^{-2}$ MeV. For $m_h$ and $v$, we have used $m_h=125.1$ GeV \cite{RPP2018} and $v=246$ GeV. 

It is expected that $\alpha^\prime$ is not as large as $\mathcal{O}\left( 1\right)$. The smaller $\lambda_{H\Phi}$ does not impose a physically meaningful constraint on $\alpha^\prime$, whose maximal value is allowed to be greater than 1. It should be noted that $\alpha^\prime$ is reported to satisfy $\alpha^\prime \lesssim 10^{-3}$ implied by astrophysical constraint on the dark radiation \cite{Ackerman2009}. If this is the case, $\Gamma\left(h \to \tilde\gamma\tilde\gamma\right)$ yields rather tiny contribution to $\Gamma\left(h \to \text{invisible}\right)$.

\subsection{Dark matter detection}
Let us discuss how our dark matter candidate satisfies the experimental requirements for relic density \cite{PLANCK2018} and for the direct search bounds \cite{XENON2018PRL}. The relic density of $\phi$ is obtained via thermal freeze out through annihilation to SM and the direct search constraint comes from the Higgs portal interaction. The dark matter annihilation into SM, $\phi^\ast \phi \rightarrow$SM+SM, via s-channel Higgs boson mediation dominates over other channels. The relic abundance of dark matter depends on the Higgs-$\phi$ coupling $\lambda_{H\Phi}$ in Eq.(\ref{Eq:VScalars}) and  mass $m_\Phi$. The direct search utilizes the scattering of $\phi$ off quarks $q$, $\phi q \rightarrow \phi q$, which occurs via the t-channel Higgs boson exchange. The dark matter - nucleon spin-independent cross section is controlled by $\lambda_{H\Phi}$ and $m_\Phi$.

\begin{figure}[t]
\begin{center}
\includegraphics{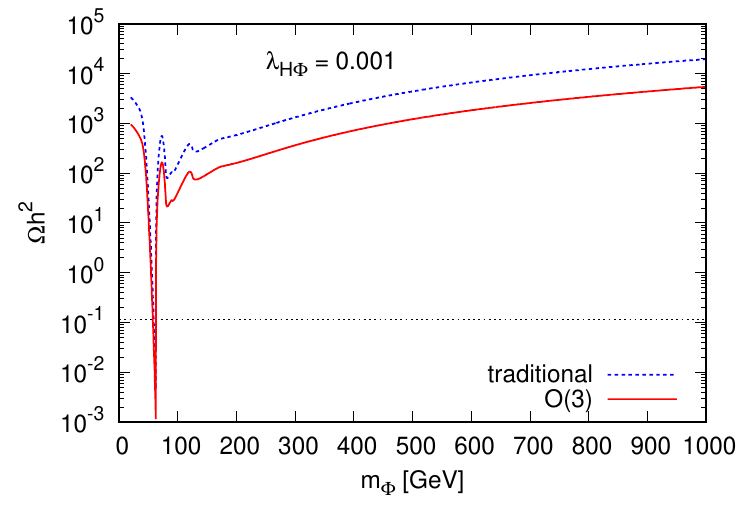} 
\includegraphics{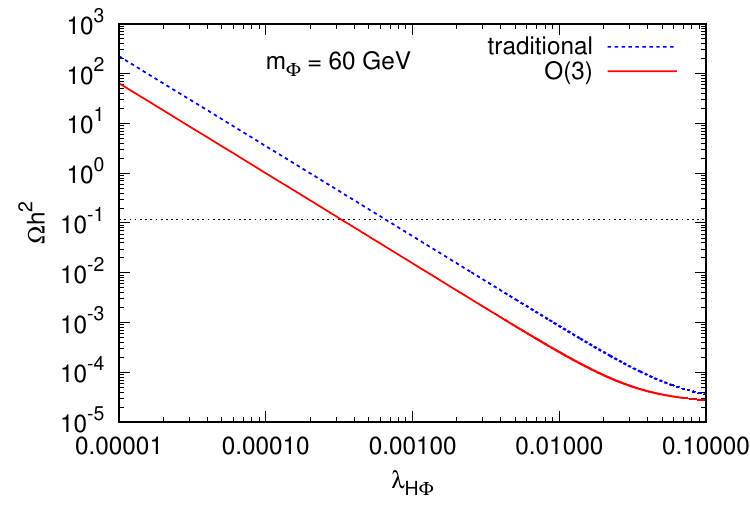}
\caption{Relic abundance of dark matter, where {\lq\lq traditional\rq\rq} and {\lq\lq O(3)\rq\rq}, respectively, stand for the singlet scalar dark matter and our $O(3)$ dark matter and where the dotted horizontal lines show the observed relic abundance $\Omega h^2 = 0.120 \pm 0.001$.}
\label{fig:omegah2}
\end{center}
\end{figure}

\begin{figure}[t]
\begin{center}
\includegraphics{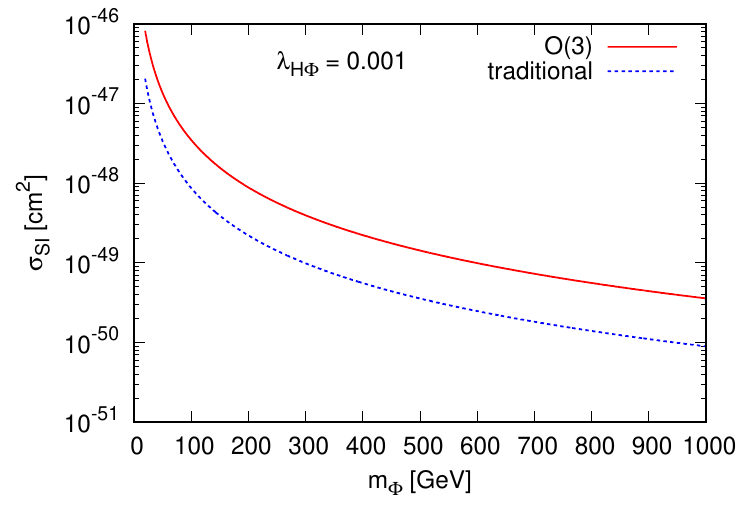}
\includegraphics{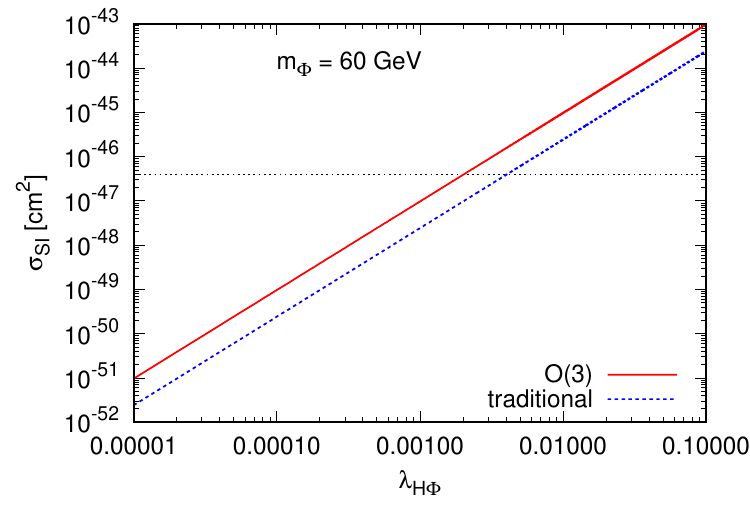}
\caption{Dark matter - nucleon spin-independent cross section, where the dotted horizontal line in the right panel shows $\sigma_{SI}= 4.1 \times 10^{-47}$ cm$^2$, above which is excluded.}
\label{fig:SI}
\end{center}
\end{figure}

Our $O(3)$ scalar dark matter subject to Eq.(\ref{Eq:GaugeLag}) is described by the same lagrangian based on the $SU(2)\times U(1)$ symmetry.  Therefore, predictions from the $O(3)$ scalar dark matter would be similar to those based on traditional scalar dark matter although every $O(3)$ dark matter particle is restricted to be neutral. As the traditional scalar dark matter, we choose singlet scalar dark matter (SDM) probably protected by a $Z_2$ parity \cite{Silveira1985,McDonald1994,Burgess2001} to compare our predictions with those of SDM, where SDM is taken to have the same mass, Higgs portal coupling and quartic coupling as those of the $O(3)$ dark matter. Our scalar dark matter has two {\lq\lq dark colors\rq\rq} since the global $O(3)$ symmetry is kept unbroken. The effect of the two {\lq\lq dark colors\rq\rq} is to enhance the annihilation rate of $O(3)$ dark matter.  As a result, the relic abundance of dark matter is less than that of SDM while the dark matter - nucleon spin-independent cross section is enhanced. Predictions from SDM are shown in FIG.\ref{fig:omegah2} and FIG.\ref{fig:SI}, where SDM is indicated by {\lq\lq traditional\rq\rq} while our $O(3)$ dark matter is indicated by {\lq\lq O(3)\rq\rq}.

We use FeynRules 2.3 \cite{FeynRules} and micrOMEGAs 5.0.9 \cite{micrOMEGAs} to compute the dark matter relic abundance and the dark matter-nucleon spin-independent scattering cross sections. We set the Higgs-$\phi$ coupling $0.00001 \le \lambda_{H\Phi} \le 0.1$ and dark matter mass $20 \le m_\Phi [\textrm{GeV}] \le 1000$. 

The relic abundance of dark matter $\Omega h^2$ is shown in FIG.\ref{fig:omegah2}. The left panel shows $\Omega h^2$-$m_\Phi$ plane for $\lambda_{H\Phi} = 0.001$ and the right panel shows $\Omega h^2$-$\lambda_{H\Phi}$ plane for $m_\Phi = 60$ GeV. The dotted horizontal lines show the observed relic abundance $\Omega h^2 = 0.120 \pm 0.001$ \cite{PLANCK2018}.  The dark matter - nucleon spin-independent cross section $\sigma_{SI}$ is shown in FIG.\ref{fig:SI}. The left panel shows $\sigma_{SI}$-$m_\Phi$ plane for $\lambda_{H\Phi} = 0.001$ and the right panel shows $\sigma_{SI}$-$\lambda_{H\Phi}$ plane for $m_\Phi = 60$ GeV. The dotted horizontal line in the right panel shows $\sigma_{SI}= 4.1 \times 10^{-47}$ cm$^2$ as the minimum of the excluded $\sigma_{SI}$ around $m_\Phi$ = 60 GeV, which is suggested by the dark matter search results given by the XENON collaboration \cite{XENON2018PRL}.  These figures correctly show the influence of the {\lq\lq dark color\rq\rq} on the $O(3)$ scalar dark matter.

\begin{figure}[h]
\begin{center}
\includegraphics{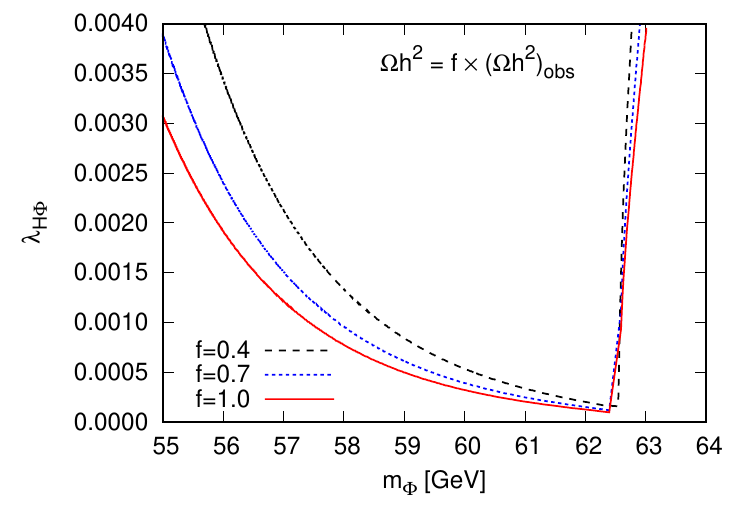}
\includegraphics{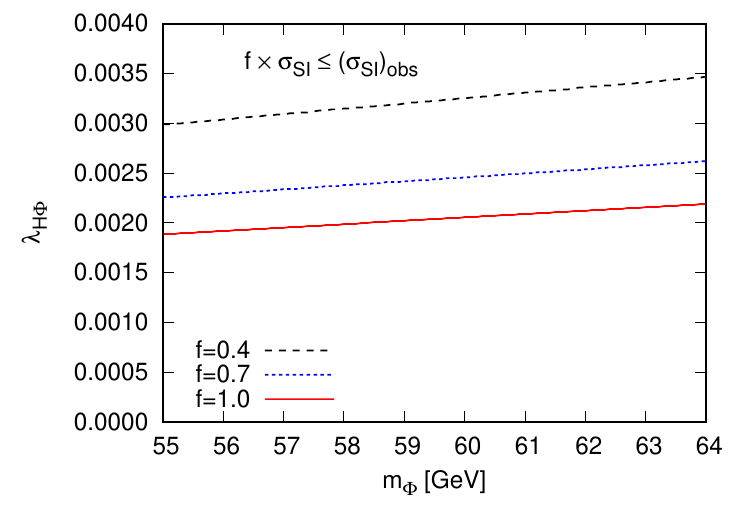}
\caption{Allowed combinations of $\lambda_{H\Phi}$ and $m_\Phi$ for $f$ = 0.4, 0.7 and 1.0 satisfying $(\Omega h^2)_{\rm obs} = 0.120 \pm 0.001$ in the left panel and $(\sigma_{SI})_{\rm obs} \le 4.1 \times 10^{-47}$ cm$^2$ in the right panel.}
\label{fig:lamHphi_m_with_fraction_omega_sigma}
\end{center}
\end{figure}

\begin{figure}[!htbp]
\begin{center}
\includegraphics{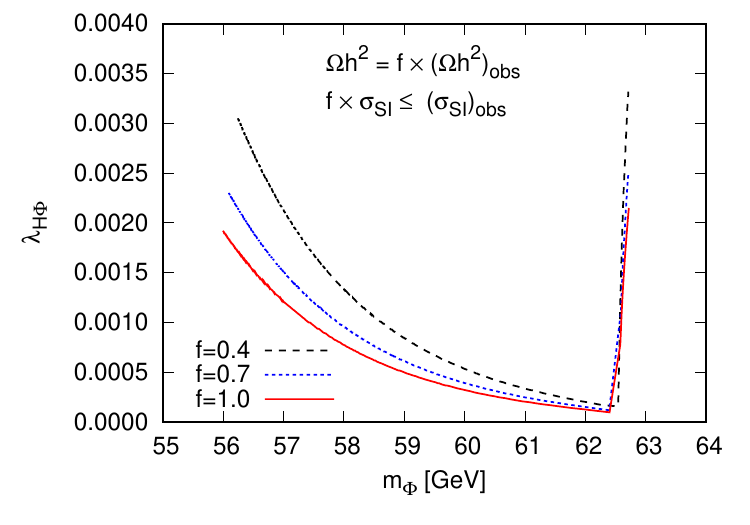}
\caption{The same as in FIG.\ref{fig:lamHphi_m_with_fraction_omega_sigma} but satisfying both constraints.}
\label{fig:lamHphi_m_with_fraction}
\end{center}
\end{figure}

If O(3) dark matter represents only a part of the total dark matter components, the O(3) dark matter signals would be normalized by a factor $f=\Omega h^2/(\Omega h^2)_{\rm obs}$ \cite{Arcadi2019arXiv}, where the subscript of {\lq\lq obs\rq\rq} stands for the observed value. In this case, the rescaled direct detection cross section $f \times \sigma_{\rm SI}$ may be comparable with the XENON limit. Considering experimentally favored regions of $\lambda_{H\Phi}$ and $m_\Phi$ indicated by both FIG.\ref{fig:omegah2} and FIG.\ref{fig:SI}, we show FIG.\ref{fig:lamHphi_m_with_fraction_omega_sigma} to provide the allowed combinations of $\lambda_{H\Phi}$ and $m_\Phi$ for $(\Omega h^2)_{\rm obs} = 0.120 \pm 0.001$ in the left panel and for $(\sigma_{SI})_{\rm obs} \le 4.1 \times 10^{-47}$ cm$^2$ in the right panel, where the kink appearing in the left panel arises around $m_\Phi = m_h/2$. Finally, including both constraints on $\Omega h^2$ and $\sigma_{\rm SI}$, we show the allowed combinations of $\lambda_{H\Phi}$ and $m_\Phi$ for $\Omega h^2 = f \times (\Omega h^2)_{\rm obs}$ and $f \times \sigma_{\rm SI} \le (\sigma_{\rm SI} )_{\rm obs}$ in FIG.{\ref{fig:lamHphi_m_with_fraction}}. We numerically find that ($0.0002 \lesssim \lambda_{H\Phi} \lesssim 0.0033$, $56.2 \lesssim m_\Phi [\textrm{GeV}] \lesssim 62.7$), ($0.0001 \lesssim \lambda_{H\Phi} \lesssim 0.0025$, $56.1 \lesssim m_\Phi [\textrm{GeV}] \lesssim 62.7$) and ($0.0001 \lesssim \lambda_{H\Phi} \lesssim 0.0022$, $56.0 \lesssim m_\Phi [\textrm{GeV}] \lesssim 62.7$), respectively, for $f$=(0.4, 0.7, 1.0) to be consistent with the observed dark matter relic abundance and the dark matter direct detection result.

\section{summary and discussions}\label{sec:summary}
Our formulation of the $O(3)$ spinor is based on the Majorana representation of the $O(4)$ spinor consisting of the $SO(3)$ spinor and its complex conjugate, The extended Pauli matrices of $O(3)$ are described by $\Gamma_i=\text{diag.}(\tau_i,\tau^\ast_i)$, which form the gamma matrices of $O(4)$ together with $\Gamma_4=-\tau_2\otimes \tau_2$,  The reflections are induced by $\Sigma_{i4}=\Sigma_i \tilde P$, where $\Sigma_i$ describe the parity flip on the specified axis as
\begin{equation}
\Sigma_1 = iD_{23}(1,\pi),
\Sigma_2 = iD_{31}(1,\pi),
\Sigma_3 = iD_{12}(1,\pi).
\end{equation}
The parity operator $\tilde P$ supports the property of $\vec{\Gamma}$ as the vector of $O(3)$ and of $\vec{\Sigma}$ as the axial vector of $O(3)$, which are, respectively, indicated by 
\begin{equation}
{\tilde P}^\dag\vec{\Gamma}{\tilde P}=-\vec{\Gamma},
\quad
{\tilde P}^\dag\vec{\Sigma}{\tilde P}=\vec{\Sigma}.
\end{equation}
Altogether, the rotation by $D_{ij}(-1,\theta)=D_{ij}(1,\theta)\Sigma_{i4}$ in the spinor space describes the corresponding $O(3)$ rotation in the vector space.  Furthermore, ${\tilde P}$ takes care of the interchange of $\phi\leftrightarrow\phi^\ast$ as $\Phi^\prime = i{\tilde P}\Phi$, which is related to the $\boldsymbol{Z}_2$ parity realized on the $O(3)$ spinor. In fact, $\tilde P$ is equivalent to the spinorial $\boldsymbol{Z}_2$ parity operator $P$: 
\begin{equation}
P=\text{diag.}\left(I, -I\right),
\end{equation}
for ($\phi$, $\phi^\ast$), which describes the $U(1)$ charge of $\Phi$.  We have also noted that $P$ as the pseudoscalar of $O(3)$ ensures that the transformed $O(3)$ spinor with the reflections included has the opposite $U(1)$ charge to the original $O(3)$ spinor

In particle physics, the $O(3)$ spinor consisting of ($\phi$, $\phi^\ast$) serve as the bosonic $O(3)$ spinor.  The fermionic spinor is composed of the Dirac spinor and its charge conjugate instead of the complex conjugate so that the internal $O(3)$ symmetry gets orthogonal to the Lorentz symmetry. The $O(3)$ symmetry gets visible when the invariance of interactions is extended to explicitly include contributions from their complex conjugates. Owing to the inherent property that both the particle and antiparticle are contained in the $O(3)$ spinor, which is constrained to be neutral, the application to particle physics is limited.  One of the viable possibilities is to regard the $O(3)$ spinors as dark matter. To see the interactions of the dark sector particles, we have constructed the lagrangians of the bosonic spinor, the fermionic spinor and the gauge bosons. The results indicate that the fermionic spinor is vectorlike and that the gauge bosons, $V_{i\mu}$ and $U_\mu$, associated with $SO(3)\times \boldsymbol{Z}_2$ behave as the axial vector and pseudoscalar, respectively. 

We have described the following scenarios of dark gauge bosons: 
\begin{enumerate}
	\item The massless dark photon is associated with the unbroken $U(1)$ symmetry generated by the spinorial $\boldsymbol{Z}_2$ parity, and $h\to{\tilde\gamma}{\tilde\gamma}$ proceeds via the one loop by $\Phi$;
	\item Three degenerated massive dark photons are associated with the spontaneously broken $O(3)/U(1)$ symmetry and $h\to{\tilde\gamma}{\tilde\gamma}$ proceeds via the one loop by $V$ with the gauge coupling to $\tilde\gamma$ as well as by $\Psi$ and $N_R$ with the $\Psi$ coupling to $\tilde\gamma$ if the decay is kinematically allowed;
	\item The massless dark photon is associated with the spontaneously broken $O(3)$ symmetry and $h\to{\tilde\gamma}{\tilde\gamma}$ proceeds via the one loop by $\tilde W^\pm$ with the gauge coupling to $\tilde \gamma$.
\end{enumerate}
We have further obtained that 
\begin{enumerate}
	\item the $\tilde \gamma$-$Z$ mixing is generated as loop corrections to give the decay of $h$ as $h\to Z\tilde \gamma$ and the kinetic $\tilde \gamma$-$\gamma$ mixing further suppressed by the $Z$ boson mass;
	\item massive dark gauge bosons decay into $\nu{\bar \nu}$;
\end{enumerate}

To speak of the feasibility of dark matter models based on the $O(3)$ spinors, we adopt the degenerated scalars of $\phi_{1,2}$ in the $O(3)$-doublet as the lightest dark matter candidate in the specific dark matter model based on the $U(1)$ gauge symmetry.  We have estimated the dark matter relic abundance and the dark matter - nucleon spin-independent cross section for dark matter direct detection experiments.  The consistent results are obtained for the Higgs-$\Phi$ coupling with $0.0001 \lesssim \lambda_{H\Phi} \lesssim 0.0022$ and the dark matter mass with $56.0 \lesssim m_\Phi [\textrm{GeV}] \lesssim 62.7$ for the case of $f(=\Omega h^2/(\Omega h^2)_{\rm obs})=1.0$.  This choice of the parameters fulfills the experimental constraint on the invisible Higgs decay.  We will present more intensive analyses \cite{O3DarkMatter}, which include $\psi$ instead of $\phi$ as the lightest dark matter candidate and other dark matter models containing the $O(3)$ dark gauge symmetry, which will be subject to constraints on the $H$-$\phi$ mixing.

Finally, influence on the neutrino oscillations can be discussed in the case of the nonvanishing $\langle \Phi\rangle$. The $O(3)$-doublet vectorlike fermion supplies four species of Majorana fermions of $\psi_{1,2}$ and $\psi^C_{1,2}$, which all couple to $N_R$ as in Eq.(\ref{Eq:RightHandedN}), and will mix with flavor neutrinos via the seesaw mechanism.  These Majorana fermions can be sterile neutrinos as dark matter \cite{Lindner2014}. To clarify detailed dark matter physics influenced by the $O(3)$-doublet vectorlike fermion as sterile neutrinos as well as to estimate their effects on neutrino oscillations is left for our future study.  Since the $CP$ transformation exchanges particles and antiparticles, which are also exchanged by $D_{23}(-1,0)$, there might be a chance to implement the $CP$ violation in the dark sector associated with the breaking of $O(3)$.

\vspace*{5mm}





\begin{thebibliography}{99}

\bibitem{Planck2016}
P.A.R. Ade et. al. (Planck Collaboration), \Journal{\APJ}{594}{A13}{2016}.

\bibitem{Arkani-Hamed2009}
See for example,
N. Arkani-Hamed, D.P. Finkbeiner, T.R. Slatyer and N. Weiner, \Journal{\PRD}{D79}{015014}{2009}.

\bibitem{Alexander2016}
J. Alexander {\it et al.}, \lq\lq Dark Sectors 2016 Workshop: Community Report\rq\rq, FERMILAB-CONF-16-421, arXiv:1608.08632 [hep-ph] (2016).

\bibitem{Arcadi2018}
For a review, see for example,
G. Arcadi, M. Dutra, P. Ghosh, M. Lindner, Y. Mambrini, M. Pierre, S. Profumo and F.S. Queiroz, \Journal{\EPJC}{78}{203}{2018} and references therein.

\bibitem{Kitabayashi2018}
T. Kitabayashi and M. Yasu\`{e}, \lq\lq Spinor Representation of $O(3)$ for $S_4$\rq\rq, arXiv:1810.02034 [hep-ph] (2018).

\bibitem{Pakvasa1979}
S. Pakvasa and H. Sugawara, \Journal{\PLBOLD}{B82}{105}{1979}.

\bibitem{Derman1979}
E. Derman and H.-S. Tsao, \Journal{\PRD}{20}{1207}{1979}.

\bibitem{Lam2008PRL}
C.S. Lam, \Journal{\PRL}{101}{121602}{2008}.

\bibitem{Bernigaud2018}
For a recent study, see for example,
F.J. de Anda and S.F. King, \Journal{\JHEP}{07}{057}{2018}.

\bibitem{Zhang2010}
H. Zhang, C.S. Li, Q.-H. Cao and Z. Li, \Journal{\PRD}{82}{075003}{2010}.

\bibitem{Minkowski1977}
P. Minkowski, \Journal{\PLBOLD}{B67}{421}{1977}.

\bibitem{Yanagida1979}
T. Yanagida, in {\it Proc. of the Workshop on Grand Unified Theory and Baryon Number of the Universe}, eds. A. Sawada and A. Sugamoto, (KEK, Japan, 1979), p.95.

\bibitem{Gell-Mann1979}
M. Gell-Mann, P. Ramond and R. Slansky in {\it Proceedings of the Supergravity Stony Brook Workshop}, eds. D. Freedman {\it et al.} (North Holland, Amsterdam, 1979).

\bibitem{Glashow1980}
S. L. Glashow in {\it Proceedings of the 1979 Carg\`{e}se Summer Institute on Quarks and Leptons}, eds. M. Levy {\it et al.} (Plenum, New York, 1980), p. 687.

\bibitem{Mohapatra1980}
R. N. Mohapatra and G. Senjanov\'{i}c, \Journal{\PRL}{44}{912}{1980}.

\bibitem{Schechter1980}
J. Schechter and J. W. F. Valle, \Journal{\PRD}{22}{2227}{1980}.

\bibitem{Dobrescu2005}
B.A. Dobrescu, \Journal{\PRL}{94}{151802}{2005}.

\bibitem{Pospelov2008}
M. Pospelov, A. Ritz and M.B. Voloshin, \Journal{\PLB}{662}{53}{2008}.

\bibitem{Hambye2009}
T. Hambye, \Journal{\JHEP}{0901}{028}{2009}.

\bibitem{Kaplan1992}
D.B. Kaplan, \Journal{\PRL}{68}{741}{1992}.

\bibitem{Dodelson1992}
S. Dodelson, B.R. Greene and L.M. Widrow, \Journal{\NPB}{372}{467}{1992}.

\bibitem{Fukugita1986}
M. Fukugita and T. Yanagida, \Journal{\PLBOLD}{172}{45}{1986}.
\bibitem{Cosme2005}
N. Cosme, L.L. Honorez and M.H.G. Tytgat, \Journal{\PRD}{72}{043505}{2005}.

\bibitem{Kaplan2009}
D.E. Kaplan, M.A. Luty and K.M. Zurek, \Journal{\PRD}{79}{115016}{2009}. 

\bibitem{Petraki2013}
K. Petraki and R.R. Volkas, \Journal{\IJMPA}{28}{1330028}{2013}.

\bibitem{Zurek2014}
K.M. Zurek, \Journal{\PREP}{537}{91}{2014}.

\bibitem{Holdom1986}
B. Holdom, \Journal{\PLBOLD}{166}{198}{1986}.

\bibitem{Biswas2015}
S. Biswas, E. Gabrielli, M. Heikinheimo and B. Mele, \Journal{\JHEP}{06}{102}{2015}.

\bibitem{Khoze2014}
V.V. Khoze and G. Ro, \Journal{\JHEP}{10}{061}{2014}.

\bibitem{Ahlers2008}
M. Ahlers, J. Jaeckel, J. Redondo and A. Ringwald, \Journal{\PRD}{78}{075005}{2008}.

\bibitem{Weinberg2013}
S. Weinberg, \Journal{\PRL}{110}{241301}{2013}.

\bibitem{Clarke2014}
D. Clarke, R.Foot and R.R. Volkas, \Journal{\JHEP}{1402}{123}{2014}.

\bibitem{Garcia2014}
C.A. Garcia Cely, \lq\lq Dark Matter Phenomenology in Scalar Extensions of the Standard Model\rq\rq. PhD thesis, Technische Universitat Munchen (TUM), 2014 (http://inspirehep.net/record/1500078).

\bibitem{Davoudiasl2013}
H. Davoudiasl and I.M. Lewis, \Journal{\PRD}{89}{055026}{2014}, 

\bibitem{Altmannshofer2015}
W. Altmannshofer, W.A. Bardeen, M. Bauer, M. Carena and J.D. Lykken, \Journal{\JHEP}{01}{032}{2015}. 

\bibitem{Dunsky2019}
D. Dunsky, L. J. Hall and K. Harigaya, \Journal{\JHEP}{07}{016}{2019}.

\bibitem{Bhattacharya2019}
See for example, S. Bhattacharya, P. Ghosh, A.K. Saha and A. Sil, \lq\lq Two component dark matter with inert Higgs doublet: neutrino mass, high scale validity and collider searches\rq\rq, arXiv:1905.12583 [hep-ph] (2019).

\bibitem{Heinemeyer2013}
S. Heinemeyer {\it et al.} (LHC Higgs Cross Section Working Group Collaboration), \lq\lq Handbook of LHC Higgs Cross Sections: 3. Higgs Properties\rq\rq, arXiv:1307.1347 [hep-ph] (2013).

\bibitem{CMSHIGGS2018}
CMS collaboration, \lq\lq Measurements of Higgs boson properties from on-shell and off-shell production in the four-lepton final state\rq\rq, CMS-PAS-HIG-18-002 (2018).

\bibitem{RPP2018}
M. Tanabashi {\it et al.} (Particle Data Group), \Journal{\PRD}{98}{030001}{2018}.

\bibitem{FeynRules}
A. Alloul, N.D. Christensen, C. Degrande, C. Duhr, and B. Fuks, \Journal{\CompPhysCom}{185}{2250}{2014}.

\bibitem{micrOMEGAs}
G. Belanger, F. Boudjema, A. Pukhov, and A. Semenov, \Journal{\CompPhysCom}{185}{960}{2014}.

\bibitem{PLANCK2018}
N. Aghanim {\it et al.} (Planck Collaboration), \lq\lq Planck 2018 results. VI. Cosmological parameters\rq\rq, arXiv:1807.06209 [astro-ph.CO] (2018).

\bibitem{Ackerman2009}
L. Ackerman, M.R. Buckley, S.M. Carroll and M. Kamionkowski, \Journal{\PRD}{79}{023519}{2009}.

\bibitem{XENON2018PRL}
E. Aprile, {\it et al.} (XENON Collaboration), \Journal{\PRL}{121}{111302}{2018}.

\bibitem{Silveira1985} 
V. Silveira and A. Zee, \Journal{\PLBOLD}{161}{136}{1985}

\bibitem{McDonald1994}
J. McDonald, \Journal{\PRD}{50}{3637}{1994}. 

\bibitem{Burgess2001}
C.P. Burgess, M. Pospelov and T. ter Veldhuis, \Journal{\NPB}{619}{709}{2001}. 

\bibitem{Arcadi2019arXiv}
See for example, G. Arcadi, A. Djouadi and M. Raidal, ``Dark Matter through the Higgs portal", arXiv:1903.03616 [hep-ph] (2019).

\bibitem{O3DarkMatter}
T. Kitabayashi and M. Yasu\`{e}, an article in preparation.

\bibitem{Lindner2014}
M. Lindner, S. Schmidt and J. Smirnov, \Journal{\JHEP}{1410}{177}{2014}. 

\end{thebibliography}



\end{document}